\documentclass[12pt]{article}
\usepackage[margin=1in]{geometry}
\usepackage{setspace}
\usepackage[latin1]{inputenc}
\usepackage{amsmath}
\usepackage{amsfonts}
\usepackage{amssymb}
\usepackage{soul}
\usepackage{bm}
\usepackage{multirow}
\usepackage{comment}
\usepackage{subcaption} 
\usepackage{tikz}
\usetikzlibrary{shapes.geometric, arrows}
\usepackage[latin1]{inputenc}
\usepackage{filecontents}
\usepackage{subcaption}
\usepackage{graphicx,hyperref}
\usepackage{natbib}
\usepackage[resetfonts]{cmap}
\usepackage{graphicx,psfrag,epsf}
\usepackage{enumerate}
\usepackage{ftnxtra}
\usepackage{array}
\hypersetup{colorlinks = true,linkcolor = blue,anchorcolor = blue,citecolor = blue,filecolor = blue,urlcolor = blue}

\usepackage{lscape}
\usepackage{float}

\begin{document}
\pagenumbering{roman}
\begin{titlepage}
	\begin{center}
		\Large \textbf{Individual-level models of disease transmission incorporating piecewise spatial risk functions }\\[0.5cm]

		Chinmoy Roy Rahul$^1$, Rob Deardon$^{1,2}$\\
		Department of Mathematics and Statistics, 
		University of Calgary$^1$\\	
  Faculty of Veterinary Medicine, University of Calgary$^2$
	\end{center}
	
	\vspace{0.7cm}

\section*{Abstract:}




Modelling epidemics is crucial for understanding the emergence, transmission, impact and control of diseases. Spatial individual-level models (ILMs) that account for population heterogeneity are a useful tool, accounting for factors such as location, vaccination status and genetic information.

Parametric forms for spatial risk functions, or kernels, are often used, but rely on strong assumptions about underlying transmission mechanisms. Here, we propose a class of non-parametric spatial disease transmission model, fitted within a Bayesian Markov chain Monte Carlo (MCMC) framework, allowing for more flexible assumptions when estimating the effect on spatial distance and infection risk.\par
We focus upon two specific forms of non-parametric spatial infection kernel: piecewise constant and piecewise linear. Although these are relatively simple forms, we find them effective. The performance of these models is examined using simulated data, including under circumstances of model misspecification, and then applied to data from the UK 2001 foot-and-mouth disease.\par

\textbf{Keywords: epidemic models; piecewise spatial risk; Bayesian Markov chain Monte Carlo; foot-and-mouth disease } 

\end{titlepage}

\pagenumbering{arabic}
\newpage

	
	\section{Introduction}\label{sec:intro}

Mathematical modeling of infectious disease provides a tool to understand the emergence of a new disease, to predict its causes, characteristics, transmission, health impact, and to project the impact of control measures (\cite{anonychuk2011health, basu2013complexity, keeling2011modeling}). Over recent years, there has been a noticeable increase in research activity on spatio-temporal models to describe infectious disease dynamics (\cite{rotejanaprasert2019spatiotemporal, o2010introduction, deardon2010inference, riley2007large}). \cite{deardon2010inference} proposed a framework of discrete-time individual-level models (ILMs) which are highly intuitive and flexible, allowing the probability of a susceptible individual being infected depending on covariates associated with both the susceptible, and surrounding infectious individuals. Note, the ``individuals'' in question may be individual people, plants or animals. However, we can also model at the level of aggregated units. The term ``individual-level'' is used since heterogeneities at whatever level we model at (for example, human, school, ranch, district, and so forth),  are being accounted for. For example, geographic location, vaccination status, or genetic information of the individual units can be accounted for.\par
Typically, a compartmental framework is used for modeling an individual's transition between disease states in infectious disease models. For example, in an $SIR$ framework, at any given time point, the susceptible ($S$) class is the group of individuals who are vulnerable to infection, the infectious ($I$) class are the potential transmitters of the infectious agent, and the removed ($R$) class represents those who have recovered from the disease, been isolated or died. \par
Inference for such models is typically carried out in a Bayesian Markov chain Monte Carlo (MCMC) framework. The Bayesian approach has firm mathematical foundations, and MCMC is a very powerful tool for carrying out inference on highly complex models (e.g., \cite{robert2004monte}).\par 
For modeling infectious disease spread, parametric models have typically been used. This generally involves some strong underlying assumptions about the mechanisms of disease spread in the population. 
However, it is often difficult to determine appropriate parametric assumptions. In addition, in the context of epidemic models, often it is not clear how best to assess the goodness of fit of parametric models, and hence difficult to measure the extent to which the assumed underlying model assumptions are in line with observed data (\cite{gardner2011goodness},  \cite{xu2016bayesian}, \cite{kypraios2018bayesian}).
This may lead us to consider using a non-parametric approach which does not assume a simple, restrictive functional form and, hence, gives us greater flexibility.\par  

In this paper, we develop a framework for simple piecewise spatial infectious disease transmission models in a Bayesian MCMC framework. This approach will help us to estimate the effect of spatial separation between individuals and the risk of infection with much more robust functional assumptions.\par

The layout of the rest of the paper is as follows. In Section \ref{method}, we introduce our proposed semi-parametric spatial model and inference method. In Section \ref{sec:simulation}, We demonstrate the simulation process and the findings in detail. In Section \ref{FMDresults}, we apply our model to $2001$ UK foot-and-mouth disease (FMD) data and exhibit the outcome. Lastly, in Section \ref{discussion}, we offer plans for further research after summing up the study.
 
\section{Methodology}\label{method}
\subsection{Individual-level Models}\label{sec:model}

\cite{deardon2010inference} defined a flexible class of models known as individual level models (ILMs), which can be used to understand the spatio-temporal dynamics of infectious disease spread in heterogeneous populations. In ILMs, individuals are considered as discrete points in time and space. Note, an event said to be occurring at discrete time $t$ represents one that is occurring within the continuous time interval $[t, t+1)$. 
\par
Here, our ILMs are placed in a discrete-time susceptible-exposed-infectious-removed (SEIR) compartmental framework, or simplifications of that framework. In the SEIR framework, an individual can be in any of the following four states at any discrete time point $t$: susceptible ($S$), exposed ($E$), infectious ($I$) or removed ($R$). Individuals who are vulnerable to the disease but are not yet infected are considered to be in the susceptible class, denoted by $S$. Individuals who are infected, but can not infect others in the population are considered to be in the exposed class, denoted by $E$. The time spent in this state is referred to as the latent period. Individuals who are infected and can infect others in the population are considered to be in the infectious class, denoted by $I$. The time spent in this state is known as the infectious period. Lastly, individuals are considered to be in the removed class, denoted by $R$, if they recover from the disease whilst acquiring immunity, die or are separated from the population (e.g., due to quarantine). An individual who is in state $S$, $E$, $I$ or $R$ at any discrete time point $t$ is said to be a member of the set $S(t)$, $E(t)$, $I(t)$ or $R(t)$, respectively. 
Here, we denote $\mu_E$ as the latent period, determining the time of transition from $E$ to $I$, and $\mu_I$ as the infectious period, determining the time of transition from $I$ to $R$. We consider $\mu_E$ and $\mu_I$ are fixed and known. However, we can extend the model where we can estimate $\mu_E$ and $\mu_I$.\par
In this paper, we consider SI (susceptible-infectious) and SEIR compartmental frameworks in simulation and real life data settings, respectively. However, the models we are concerned with can also be fitted within other frameworks (e.g., SIR (susceptible-infectious-removed)).\par

The following equation gives the general form of the epidemic ILM, introduced in \cite{deardon2010inference}, which describes the probability $P(i,t)$ of a susceptible individual $i$ being infected within the continuous time interval $[t, t+1)$; $i.e.$, at discrete time point $t$:
\begin{align}\label{meth.sec.2.2.eq1}
    P(i,t) = 1 -\exp{\Big[\Big\{-\Omega_S(i) \sum_{j\in I(t)}\Omega_T(j)k(i,j)\Big\} - \epsilon(i,t)\Big]}; \Omega_S(i), \Omega_T(j), k(i,j), \epsilon(i,t) \ge 0
\end{align}
where: $I(t)$ is the set of infectious individuals at time $t$; $\Omega_S(i)$ is a susceptibility function representing risk factors associated with susceptible $i$ contracting the disease, and $\Omega_T(j)$ is a transmissibility function representing risk factors associated with infectious individual $j$ passing the disease on. For example, $\Omega_S(i)$ and $\Omega_T(j)$ may include environmental or genetic risk factors associated with individuals $i$ or $j$, respectively. The infection kernel $k(i,j)$ represents risk factors shared between pairs of infectious and susceptible individuals. The infection kernel $k(i,j)$ could be a spatial function of Euclidean or road distance (\cite{savill2006topographic}). Otherwise, it could represent connections in the population via a network. When the infection kernel is based upon spatial distance, we often refer to it as a spatial kernel. Infections that are not well explained by the functions $\Omega_s(i)$, $\Omega_T(j)$, or $k(i,j)$, can be introduced by the so-called sparks term, $\epsilon(i,t)$. This might account for infections originating from outside the observed population. \par

\subsection{Non-parametric Spatial ILMs}\label{sec:model2}

We now turn to spatial forms of the model, in which we ignore any covariate and, initially, sparks effect. 
A typical parametric form of a spatial ILM, using a power law distance kernel, consists of setting: $\Omega_S(i) =\alpha > 0$, where $\alpha$ is a baseline susceptibility parameter; $\Omega_T(j) = 1$; $\epsilon(i,t) = 0$; and $k(i,j) = d_{ij}^{-\beta}$, where $d_{ij}$ is the distance between individuals $i$ and $j$, and $\beta >0$ is a spatial parameter. This gives the model: 

\begin{align}\label{powerlawILM}
P_{it}(\theta) = 1 - \exp{\{-\alpha \sum_{j \in I_t} d_{ij}^{-\beta}\}}, ~~ \beta > 0
\end{align}

\noindent Here, we consider a non-parametric form of the infection kernel, giving us a new non-parametric ILM framework, in which we set $\Omega_S(i) = \Omega_T(j) = 1$, $\epsilon(i,t) = 0$, and $k(i,j) = \Tilde{k}(i,j)$ to give 
\begin{align}\label{meth.sec.2.3.eq4}
    P_{it}(\theta) = 1 - \exp{\{-\sum_{j \in I_t}\tilde{k}(i,j)\}}
\end{align}
where, 
\begin{equation}\label{meth.sec.2.3.equ5}
  \tilde{k}(i,j) =
    \begin{cases}
      \tilde{k}_l(d_{ij}) & \delta_{l-1} \le d_{ij} < \delta_l ; l = 1,\ldots, n-1\\
      \tilde{k}_{l}(d_{ij}) & d_{ij} \geq \delta_{l-1}; l = n,
    \end{cases}       
\end{equation}
and where, $\delta_0,\delta_1,\ldots,\delta_{n-1}$ are change points and $\delta_0 = 0$. 

In turn, we consider two specific forms of this model. First,  we let $\tilde{k}_l(d_{ij}) = \alpha_l$ where $\alpha_l \in \mathbb{R}^+$, to define an {\bf n-step piecewise constant model}, with infection kernel:

\begin{equation}\label{meth.sec.2.3.equ7_three}
  \tilde{k}(i,j) =
    \begin{cases}
      \alpha_1 & \delta_0 \le d_{ij} < \delta_1,\\
      \alpha_2 & \delta_1 \le d_{ij} < \delta_2,\\
      & ~~~~~~~~ \vdots \\
      \alpha_{n-1} & \delta_{n-2} \le d_{ij} < \delta_{n-1},\\
      \alpha_n & d_{ij} \geq \delta_{n-1}.
    \end{cases}       
\end{equation}

\noindent Second, if we let $\tilde{k}_l(d_{ij}) = \alpha_l + \beta_l d_{ij}$ where $\alpha_l \in \mathbb{R}^+$ and $\beta_l \in \mathbb{R}^-$, then we have an {\bf n-step piecewise linear model}, with infection kernel:

\begin{equation}\label{meth.sec.2.3.equ52_linear}
  \tilde{k}(i,j) =
    \begin{cases}
    \alpha_1 + \beta_1 d_{ij} & \delta_0 \le d_{ij} < \delta_1,\\
    \alpha_2 + \beta_2 d_{ij} & \delta_1 \le d_{ij} < \delta_2,\\
      & ~~~~~~~~ \vdots \\
    \alpha_{n-1} + \beta_{n-1} d_{ij} & \delta_{n-2} \le d_{ij} < \delta_{n-1}\\
    \alpha_n + \beta_n d_{ij} & d_{ij} \geq \delta_{n-1},
    \end{cases}       
\end{equation}

Note that, an n-step piecewise kernel has n-1 change points.
In this study, we consider both scenarios where the change points are fixed before fitting the model, or considered unknown to some degree and thus estimated within the ILM inducing more flexibility.

\subsection{Bayesian Inference}\label{sec.bayesian}
Here, inference is carried out within a Bayesian framework, using a random-walk Metropolis-Hastings algorithm  (\cite{metropolis1953equation}, \cite{hastings1970monte}, \cite{chib1995understanding}). 
With information about which infection events occur at discrete time points $t = 1,\ldots,(T-1)$, the likelihood can be easily defined by the product of the probability of infection and non-infection events over those time points. For an SEIR ILM, the form of the likelihood is

\begin{align}\label{meth.sec.2.2.eq2}
    f(X|\theta) = \prod_{t = 1}^{T-1}f_t(X|\theta),
\end{align}
where,

\begin{align}\label{meth.sec.2.2.eq3}
    f_t(X|\theta) = \left[\prod_{i\in E(t+1)\backslash E(t)} P(i,t)\right] \left[\prod_{i\in S(t+1)} \{1 - P(i,t)\}\right],
\end{align}
and where, $\theta$ is the unknown parameters vector,  $X$ is the observed epidemic data and $E(t+1)\backslash E(t)$ is the set of newly exposed individuals at time $t+1$. Typically, the observed epidemic data contains location, individual-level event times (e.g., infection time, removal time) and any relevant individual-level covariates.\par

Under an SI or SIR framework we replace equation (\ref{meth.sec.2.2.eq3}) with

\begin{align}\label{meth.sec.2.2.eq4}
    f_t(X|\theta) = \left[\prod_{i\in I(t+1)\backslash I(t)} P(i,t)\right] \left[\prod_{i\in S(t+1)} \{1 - P(i,t)\}\right],
\end{align}
where $I(t+1)\backslash I(t)$ is the set of newly infectious or infected farms at time $t+1$.

When change points are fixed, we assume priors for our parameters $\theta = (\alpha_1, \alpha_2, \ldots, \alpha_n)$ for piecewise constant kernel and $\theta = ((\alpha_1,\beta_1), (\alpha_2,\beta_2), \ldots, (\alpha_n,\beta_n))$ for piecewise linear kernel, respectively, and combine the respective prior with the likelihood as given in equation (\ref{meth.sec.2.2.eq2}) to form the posterior distribution $\pi(\theta|X)$ up to a constant of proportionality. However, when change points are unknown, we assume priors for our parameters $\theta = (\alpha_1, \alpha_2, \ldots, \alpha_n,
\delta_1,\delta_2,\ldots,\delta_{n-1})$ for piecewise constant kernel and $\theta = ((\alpha_1,\beta_1), (\alpha_2,\beta_2), \ldots, \\(\alpha_n,\beta_n),\delta_1,\delta_2,\ldots,\delta_{n-1})$ for piecewise linear 
 kernel, respectively. Note, we assume $\delta_0 = 0$ is fixed.

\section{Simulation Results}\label{sec:simulation}

In this section, we explain the simulation study, in an SI compartmental framework, used to investigate the performance of the proposed non-parametric spatial ILM regarding its ability to determine the infectious disease dynamics. \par

\subsection{Constant piecewise Kernel}
We consider a population of $N = 400$ individuals with spatial coordinates  $\underline{x} = (x_1,\ldots,x_{400})$ and $\underline{y}= (y_1,\ldots,y_{400})$, generated from independent uniform distributions  $ x \sim U[0,A]$ and  $ y \sim U[0,A]$. Here, we consider the population with either $A = 10$ or $A = 25$ in which an infectious disease can spread. The piecewise constant kernels (equations (\ref{meth.sec.2.3.eq4}) and (\ref{meth.sec.2.3.equ7_three})) are used for simulating epidemics. For models with fixed change points ($\delta$'s), we consider two-step (n = 2), three-step (n = 3), seven-step (n = 7) and ten-step (n = 10) cases, and for unknown change points, we  consider two-step and three-step cases. The parameters used in these different scenarios are given in Figures \ref{fig:2_f_graphs}, \ref{fig:2_3_graphs}, and Figures \ref{fig:7_f_graphs}, \ref{fig:10_f_graphs} (see Appendix), respectively. We used ``vague" positive half-normal priors for the $\alpha$'s ($\alpha_l \sim N^{+}(0,10^5)$; $l = 1,2,\ldots, n$) for fixed and unknown change points. Under scenarios in which the change points are unknown, for two-step cases, we used ``vague" prior $\delta_1 \sim U(0,10)$, and for the three-step cases, we used ``weakly" informative priors $\delta_1 \sim U(0,2)$ and $\delta_2 \sim U(2,4)$. In this case, we use weakly informative priors because utilizing vague priors results in identifiability problems and poor MCMC mixing. In theory, we can use higher number of steps when the change points are unknown. However, for this size of population and epidemic dynamics, we find it increasingly difficult to obtain stationary MCMC algorithms with a larger number of unknown change points. Epidemics are simulated for $T= 20$ time units. We assess MCMC convergence by visually inspecting traceplots and utilizing Geweke's diagnostic (\cite{geweke1992evaluating}) for each model parameter. If any chains fail to converge according to Geweke's diagnostic test, we then apply the Gelman-Rubin diagnostic (\cite{gelman1992inference}) using three MCMC chains. Moreover, we calculate the posterior mean, median, and $95\%$ percentile intervals (PIs) for each model parameters. The Julia language version $1.6.3.$ (\cite{bezanson2017julia}) run on the University of Calgary's Advanced Research Computing (ARC) network was used to carry out the MCMC and epidemic simulation. Plotting was performed in RStudio and Jupyter Notebook version $6.0.3$.\par


\begin{figure}[h!tpb]
     \centering
     \begin{subfigure}[b]{1\textwidth}
         \centering
         \includegraphics[width=1 \textwidth]{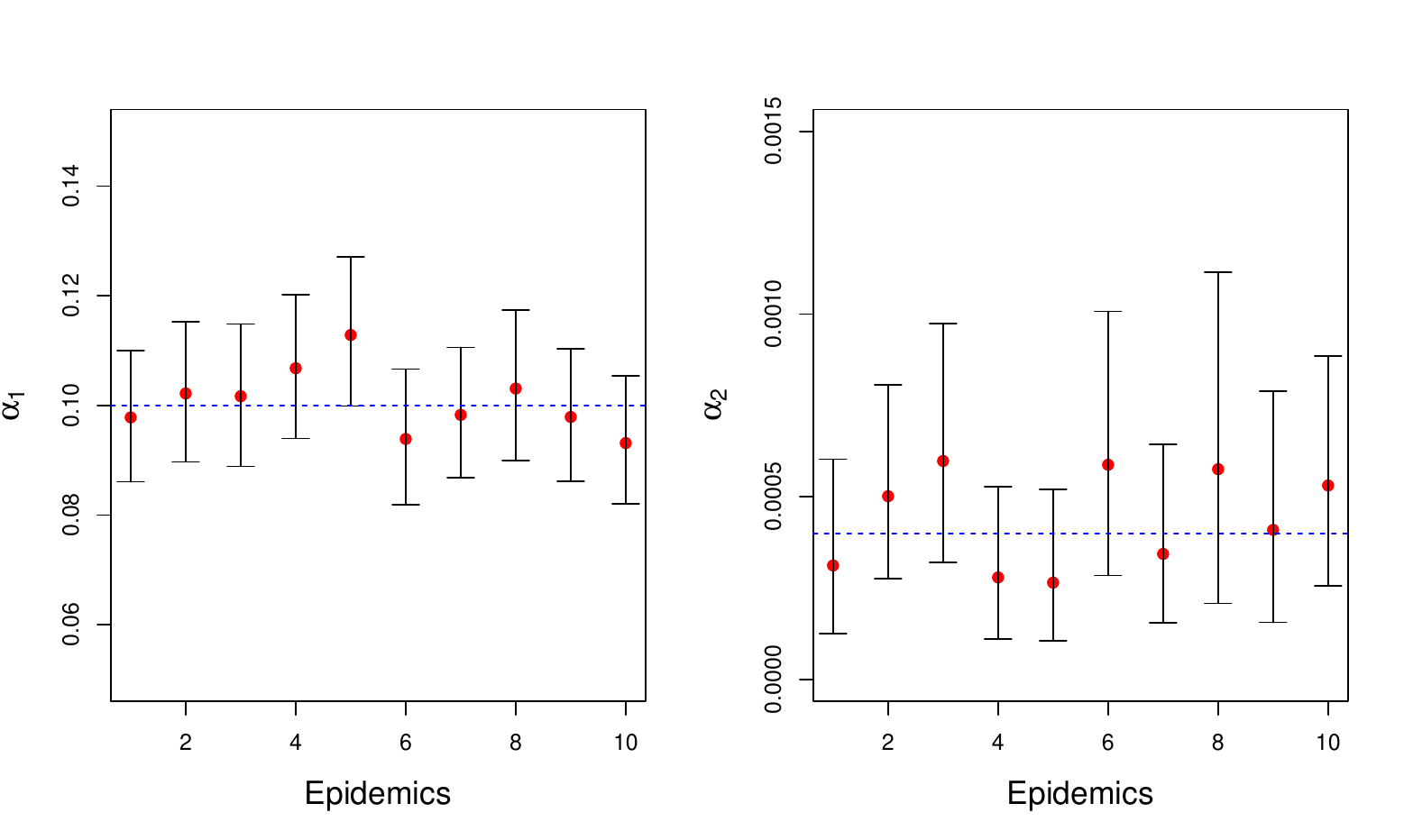}
         \caption{}
         \label{fig:y equals x}
     \end{subfigure}
     \vfill
     \begin{subfigure}[b]{1\textwidth}
         \centering
     \includegraphics[width=1\textwidth]{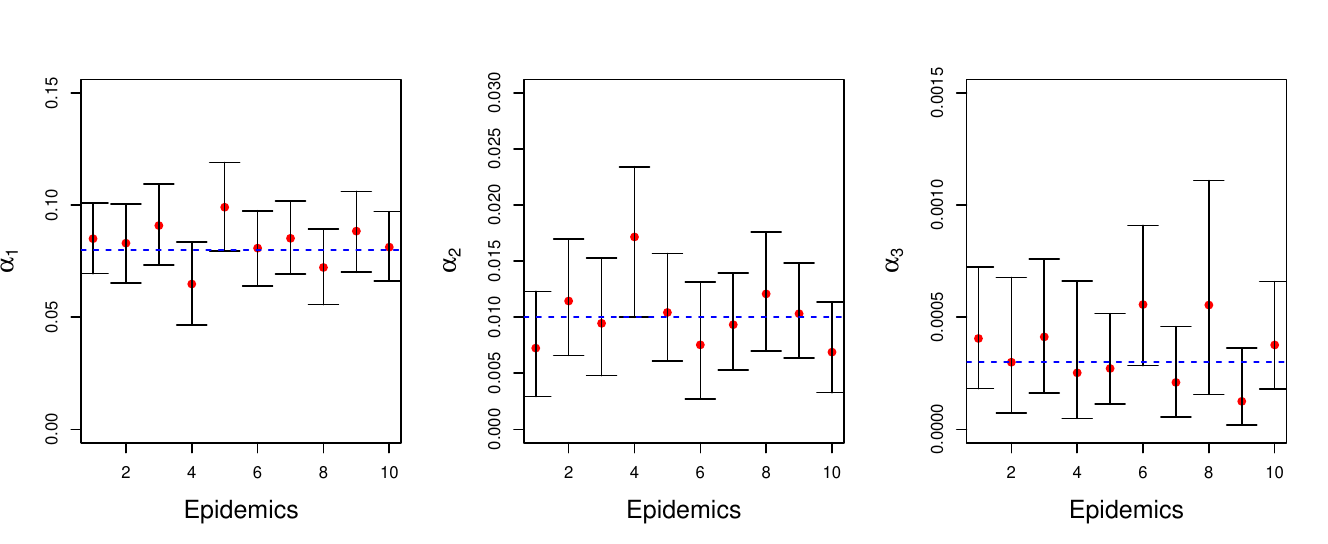}
         \caption{}
         \label{fig:three sin x}
     \end{subfigure}
     \caption{\small{Posterior medians (red points) and $95\%$ percentile intervals (PIs) for $\alpha_1,\alpha_2,\alpha_3$ for $10$ different simulated epidemics when the change points ($\delta$'s) are considered fixed. The true parameter values (a) $\alpha_1 = 0.10$, and $\alpha_2 = 0.0004$ for two-step cases with fixed $\delta_1 = 2$, and (b) $\alpha_1 = 0.08 ,\alpha_2 = 0.01$, and $\alpha_3 = 0.0003$ for three-step cases with fixed $\delta_1 = 1.5$, and $\delta_2 = 3$ are represented by the blue dashed line. } }\label{fig:2_f_graphs}
\end{figure}

\begin{figure}[h!tpb]
     \centering
     \begin{subfigure}[b]{1.0\textwidth}
         \centering
        \includegraphics[width=\textwidth,height=140 pt]{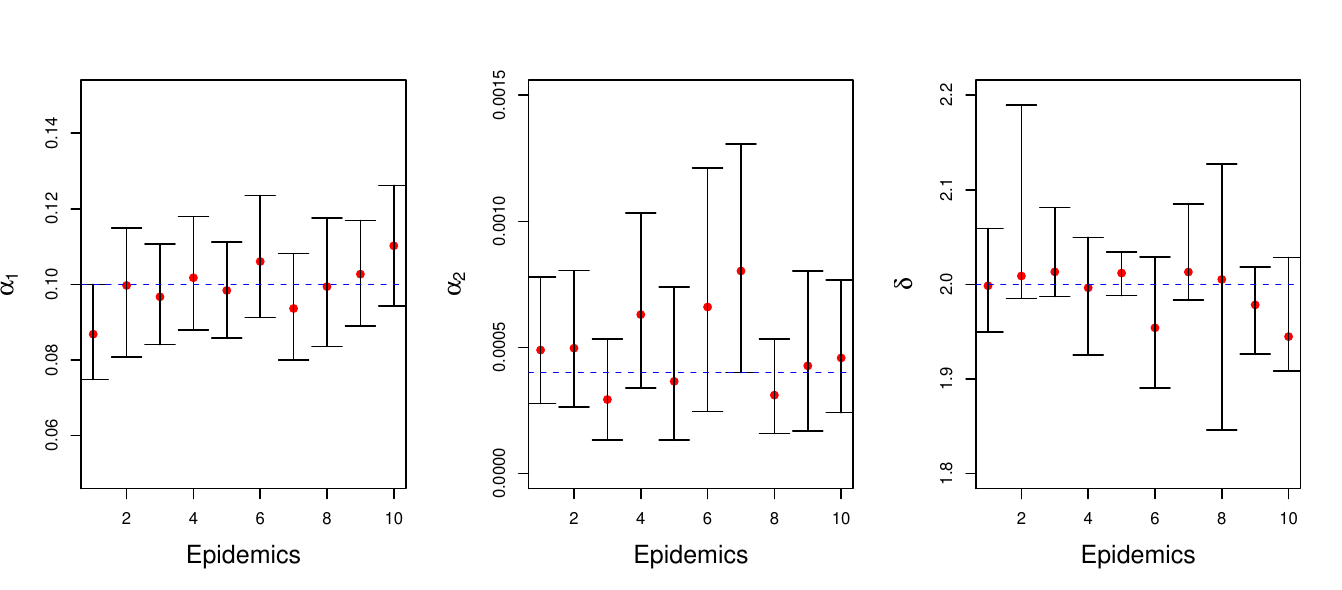}
         \caption{}
         \label{fig:y equals x}
     \end{subfigure}
     \vfill
     \begin{subfigure}[b]{1\textwidth}
         \centering
        \includegraphics[width=\textwidth,height=270 pt]{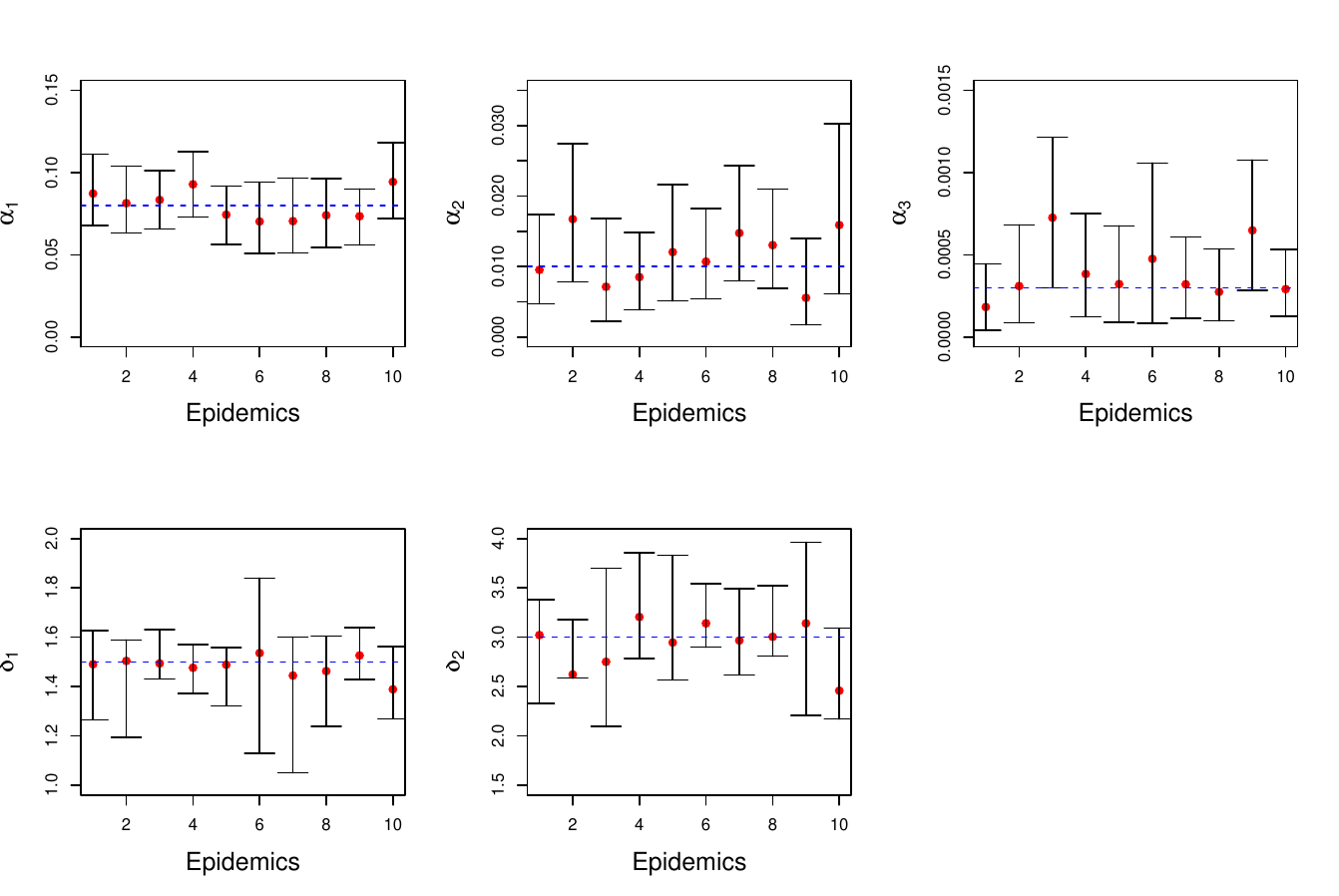}
         \caption{}
         \label{fig:three sin x}
     \end{subfigure}
     \caption{\small{Posterior medians (red points) and $95\%$ PIs for $\alpha_1,\alpha_2,\alpha_3,
     \delta_1, \delta_2$ for $10$ different simulated epidemics when the change points ($\delta$'s) are estimated. The true parameter values: (a) $\alpha_1 = 0.10 ,\alpha_2 = 0.0004$, and $\delta_1 = 2$ for two-step cases (b) $\alpha_1 = 0.08 ,\alpha_2 = 0.01, \alpha_3 = 0.0003, \delta_1 = 1.5$, and $\delta_2 = 3.0$ for three-step cases are represented by the blue dashed line.} }\label{fig:2_3_graphs}
\end{figure}

Figure \ref{fig:2_f_graphs} shows the posterior medians and $95\%$ PIs for the piecewise constant model  parameters for ten different epidemics and populations when the change points are considered fixed under two-step and three-step cases. Under two-step and three-step cases, the posterior medians of all the parameters are close to their true values, and all $95\%$ percentile intervals (PIs) capture the true parameter values that were used to generate the epidemics.\par
Similarly, Figure \ref{fig:7_f_graphs} and Figure \ref{fig:10_f_graphs} (see Appendix) show the posterior medians and $95\%$ PIs for the constant piecewise ILM parameters for ten epidemics when the change points were considered fixed, under seven-step and ten-step cases, respectively. The seven-step and ten-step cases show how our model works for a higher number of steps. Once again, the true parameter values used to generate data for both of these cases with fixed change points fall within their posterior $95\%$ PIs, with posterior medians generally close to the true parameter values.\par
Posterior medians and $95\%$ PIs of the parameters for unknown change points under two-step and three-step cases for ten simulated epidemics, are shown in Figure \ref{fig:2_3_graphs}. We can see that, under both cases, the posterior median is close to its true parameter value with varying levels of posterior uncertainty.\par 
The posterior estimate from all the different cases of piecewise constant kernel for fixed and unknown change points suggest that the proposed model with  piecewise constant kernel in Section \ref{sec:model2} is capable of capturing appropriate estimates if the generating model is fitted to simulated data.

\subsubsection*{Misspecified Model Example}\label{parametric_example1}
We now consider fitting various piecewise constant kernel ILMs when the data is actually generated by a simple parametric spatial ILM. The goal here is to assess the performance in terms of model fit and robustness to choices about numbers and positions of change points. Epidemics are generated from a parametric model with a power-law kernel, across population of size N = 250 located at positions $(x_1,y_1)\ldots(x_{250},y_{250})$, generated from independent uniform distributions $ x \sim U[0,50]$ and $y \sim U[0,50]$. The specific epidemic generating model is given by equation (\ref{powerlawILM}). For illustration, we show results for one typical epidemic (Figure \ref{epi_curve}, see Appendix) of those generated from this model using the parameter values $\alpha = 0.30$ and $\beta = 2$. Note, these values follow those used in \cite{kwong2012linearized}. We fit a variety of constant piecewise kernel ILMs to this data. For change point models with fixed $\delta$ values, we consider two-step, three-step, four-step and five-step cases, and for unknown $\delta$ values we consider two-step and three-step cases. We used ``vague" positive half-normal priors for the $\alpha$'s (i.e., $\alpha_l \sim N^{+}(0,10^5)$; $l = 1,2,\ldots, n$) for fixed and unknown change points. Once again, when the change points are unknown, for the two-step case, we use ``vague" prior $\delta_1 \sim U(0,10)$, and for the three-step case we use ``weakly" informative priors $\delta_1 \sim U(1.5,4)$ and $\delta_2 \sim U(4,8)$, respectively. Once more, in this case, weakly informative priors are chosen because employing a vague prior leads to non-identifiability.
Epidemics are simulated for $T= 49$ time units. MCMC convergence is assessed as described previously.
The posterior mean, median and $95\%$ PIs for the model parameters and the posterior predictive distribution of the epidemic curve are estimated. The epidemic curve refers to the number of new infections at each time point (i.e., incidence curve). We also consider choosing between fitted models with differing change points $\delta$s and/or  number of change points $\delta$s using the Deviance Information Criteria (DIC) (\cite{spiegelhalter2014deviance}).\par
The estimated posterior median (and $95\%$ PIs) when fitting the true parametric model parameters to the data set generated from the same kernel, were $\alpha = 0.32$ $(0.24, 0.43)$ and $\beta = 2.01$ $(1.85, 2.18)$. The DIC is $1257$ for this parametric model for this particular data.\par
Figure \ref{fig:para_fix} (see Appendix) shows the true kernel and fitted piecewise constant kernel models for different fixed change points under the posterior median and $95\%$ PIs. Figure \ref{fig:para_fix}(a) shows that the estimate of the kernel under the true power-law model is very good, as we can expect. We can also see that the change point models, in Figures \ref{fig:para_fix}(b)-(j), fitted offer as a reasonably good approximation to the true model under the posterior median. In addition, the fit of the estimated kernel tends to improve as the number of change points increases. It would appear that the piecewise constant kernel with appropriate number of fixed change points with significant size, the $95\%$ PIs, can approximate the parametric kernel reasonably well.\par
Table \ref{tab:parametric_fixed} shows that the DIC values of the piecewise constant model with different change points are higher compared to that of the parametric model, implying a worse fit, as we would expect. However, the DIC value for the ILM with four change points ($\delta_1 = 2, \delta_2 = 4, \delta_3 = 6$ and $\delta_4 = 8$) is smaller then all other piecewise constant models suggesting it is the ``best" in approximating the power-law kernel. \par
Results for the constant piecewise kernel ILMs with unknown change points are shown in Figure \ref{fig:para_varying} (see Appendix). We can see that ILMs with two change points (three-step case) tend to give a better approximation to the true kernel. This is backed up by the fact that the DIC value of the two change point ILMs is lower than that of the one change point model (Table \ref{tab:parametric_varying}).\par
The posterior predictive distribution of the epidemic curve from one typical epidemic simulation under true parametric model and other constant piecewise models for fixed and unknown change points are shown in Figures \ref{fig:1_fixed_PI} and \ref{fig:2_varying_PI} (see Appendix). We can see that the true epidemic curve is very well enveloped by the $95\%$ PIs of the epidemic curve under the true power-law kernel model. However, we also see similarly positive results for the various change point models, whether the change points are fixed or considered unknown (Figures \ref{fig:1_fixed_PI}(b-j) and Figures \ref{fig:2_varying_PI}(b-c)). This suggests that epidemic simulation is reasonably robust to the exact choice of piecewise constant kernel used, and that our approach is flexible enough to capture epidemic dynamics when the true kernel is based on a power-law.\par

 \begin{table}[h]
     \centering
\small
\begin{tabular}{ | >{\centering\arraybackslash}m{2.0cm} |>{\centering\arraybackslash}m{2.cm} |>{\centering\arraybackslash}m{3.0cm} | }
  
  \hline
  Model &   Fixed $\delta$ & DIC \\
  \hline
  Parametric & & 1257 \\
  \hline
  Two-step  & 6 &  1355 \\
  \hline
  Two-step & 10 & 1352  \\
  \hline
  Two-step & 15 & 1360  \\
  \hline
  Three-step & (8,16) & 1353  \\
  \hline
  Three-step  & (10,20) & 1352  \\
  \hline
  Four-step & (2,4,6) & 1320  \\
   \hline
  Four-step & (6,12,18) & 1348  \\
  \hline
  Five-step & (2,4,6,8) & 1316  \\
   \hline
  Five-step & (5,10,15,20) & 1340  \\
  \hline
  \end{tabular}
     \caption{DIC values for power-law model and different semi-parametric ILMs with piecewise constant kernel when the change point ($\delta$) is fixed.}
     \label{tab:parametric_fixed}
 
    \end{table}

\begin{table}[h]
     \centering
\small
\begin{tabular}{ | >{\centering\arraybackslash}m{2.0cm} |>{\centering\arraybackslash}m{5.0cm} |
>{\centering\arraybackslash}m{3.0cm} | }
  
  \hline
  Model &   Estimated  $\delta$ ($95\% PI$) & DIC \\
  \hline
  Parametric & & 1257 \\
  \hline
  Two-step & $\delta_1 = $2.78(2.70,3.67) &  1336 \\
  \hline
  Three-step & $\delta_1 = $1.83(1.52,2.27), $\delta_2 = $6.61(5.41,7.91) & 1304  \\
  \hline
  
  \end{tabular}
     \caption{DIC values for power-law model and different semi-parametric ILMs with piecewise constant kernel when the change point ($\delta$) is estimated. }
     \label{tab:parametric_varying}
 
    \end{table}

\subsection{Linear Piecewise Kernel} 
Here, we consider a new population of $N = 250$ individuals with spatial coordinates  $\underline{x} = (x_1,\ldots,x_{250})$ and $\underline{y}= (y_1,\ldots,y_{250})$, generated from independent uniform distributions $x\sim U[0,50]$ and $y\sim U[0,50]$ in which an infectious disease can spread. The piecewise linear kernels (equations (\ref{meth.sec.2.3.eq4}) and (\ref{meth.sec.2.3.equ52_linear})) are used for simulating epidemics. We present results from one, two, and three change point linear kernel models with fixed change points ($\delta$'s), and from one and two change point linear kernel models with unknown change points. The parameters used in these different models are given in Figure \ref{fig:L_f_graphs}, and Figures \ref{fig:L_f3_graphs} and \ref{fig:Lv_graphs} (see Appendix), respectively. We used ``vague" positive half-normal priors for the $\alpha$'s ($\alpha_l \sim N^{+}(0,10^5)$; $l = 1,2,\ldots, n$) and $\beta$'s ($\beta_l \sim N^{-}(0,10^5)$; $l = 1,2,\ldots, n$) for fixed and unknown change points.
Under scenarios in which the change points are unknown, for one change point model, we used a ``vague" prior $\delta_1 \sim U(0,10)$, and for the two change point model, we used ``weakly" informative priors $\delta_1 \sim U(0,3)$ and $\delta_2 \sim U(3,5)$. We use weakly informative priors because the use of a vague prior leads to partial-identifiability, as described previously.\par 
We also consider the ``smoothing prior" used by \cite{kwong2012linearized}. This prior can be used in a piecewise linear kernel model to ``force" the linear pieces of the kernel to meet approximately at the change points. Specifically, the smoothing prior is placed on the difference between the kernel levels given under each linear piece at the change points, $d_l = (\alpha_{l-1} - \alpha_l) + \delta_{l-1}(\beta_{l-1} - \beta_l)$, where $l = 2,3,\ldots, n$. Specifically, we let $d_l \sim Exp(D)$, where $\mathbb{E}(d_l) = D$ is the scale parameter.\par

Epidemics are simulated for $T= 20$ time units. 
Where strong posterior correlation between the $\alpha_l$ and $\beta_l$ parameters for $\delta_l = 1,2,\ldots, n$ is observed then each pair ($\alpha_l, \beta_l$) is updated via a bivariate Gaussian random walk proposal in the Metropolis Hastings algorithm. We estimate the posterior covariance by using  single-parameter Gaussian random walk updates in an initial MCMC chain. A scaled version of this covariance estimate is then used as the covariance in the bivariate proposal in a new MCMC algorithm. \par
Figure \ref{fig:L_f_graphs} shows the posterior medians and $95\%$ PIs for the linear kernel parameters in model (\ref{meth.sec.2.3.equ52_linear}) for ten different epidemics. Under one and two change point models, the posterior median estimates of all the parameters are reasonably close to their true values and all $95\%$ PIs capture the true parameter values.\par
Similarly, Figure \ref{fig:L_f3_graphs} (see Appendix) shows the posterior medians and $95\%$ PIs for the linear kernel parameters for ten epidemics when the change points were considered fixed, under the four change point model. The true parameter values are again well estimated.\par
Posterior medians and $95\%$ PIs of the linear kernel parameters for unknown change points under one and two change point models are shown in Figure \ref{fig:Lv_graphs} (see Appendix). Once again, the parameters of the kernel are well estimated.\par 
The posterior estimates from all the different cases of the piecewise linear  kernel for fixed and unknown change points suggest that the ILM with a piecewise linear spacial kernel is capable of capturing appropriate estimates if the correct model is fitted to epidemic data.\par

\begin{figure}[h!tpb]
     \centering
     \begin{subfigure}[b]{1\textwidth}
         \centering
         \includegraphics[width=0.78\textwidth]{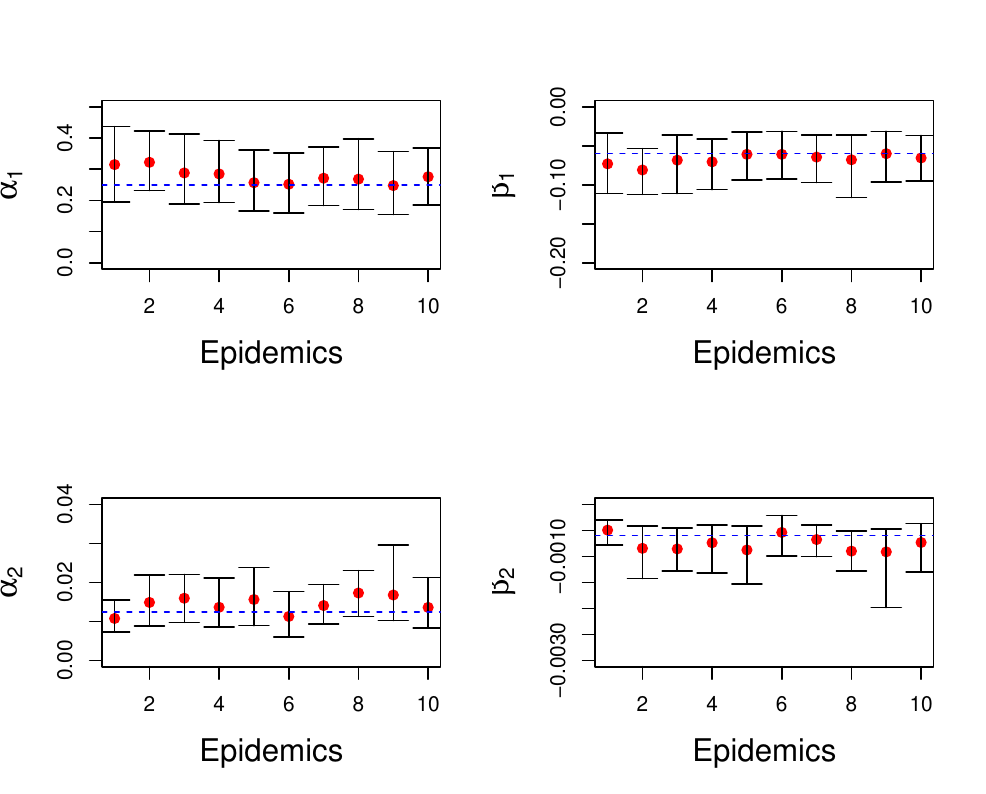}
         \caption{}
         \label{fig:y equals x}
     \end{subfigure}
     \vfill
     \begin{subfigure}[b]{1\textwidth}
         \centering
     \includegraphics[width=0.85\textwidth]{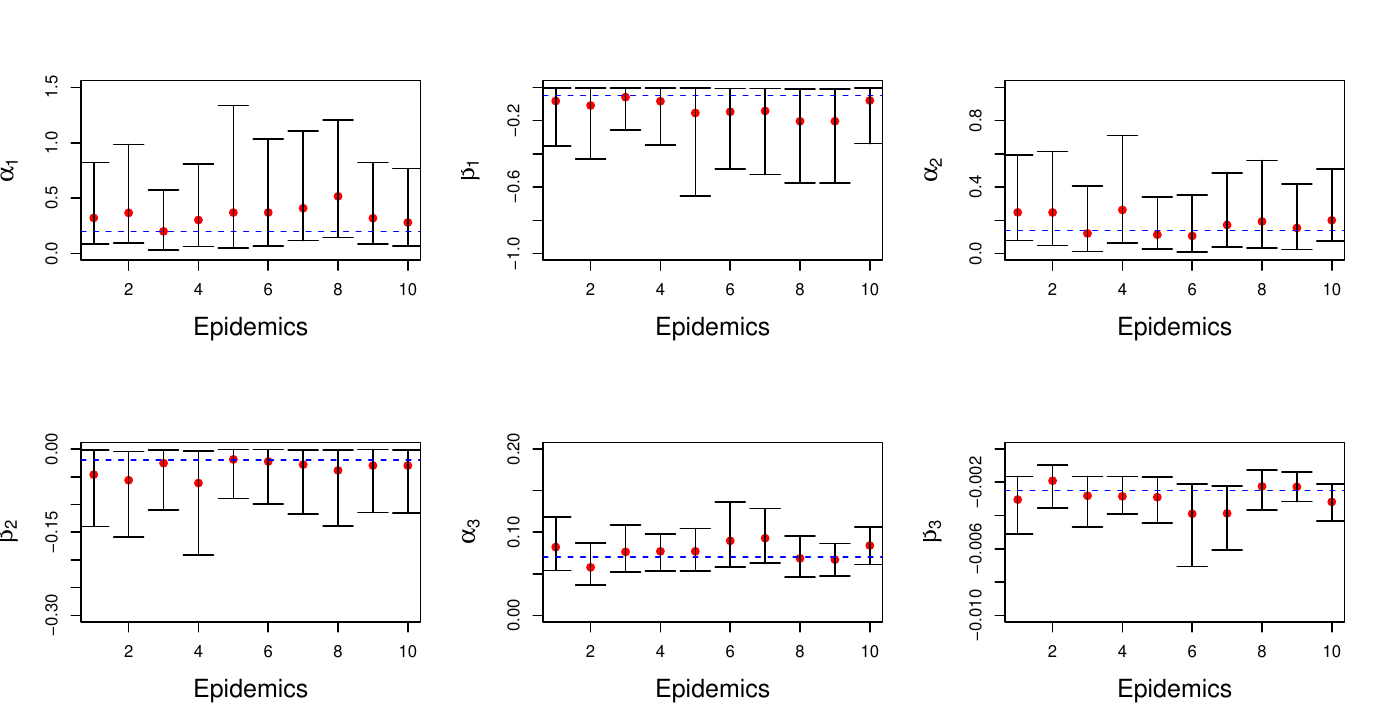}
         \caption{}
         \label{fig:three sin x}
     \end{subfigure}
     \caption{\small{Posterior medians (red points) and $95\%$ PIs for $\alpha_i,\beta_i$ where $i = 1,2,3$ for $10$ different simulated epidemics when the change points ($\delta$'s) are considered fixed. The true parameter values (a) $\alpha_1 = 0.25, \beta_1 = -0.06, \alpha_2 = 0.0124, \beta_2 = -0.0006$ for one change point linear kernel models with fixed $\delta_1 = 4$, and (b) $\alpha_1 = 0.20, \beta_1 = -0.05, \alpha_2 = 0.14, \beta_2 = -0.02, \alpha_3 = 0.07, \beta_3 = -0.0025$  for two change points linear kernel models with fixed $\delta_1 = 2$ and $\delta_2 = 4$ are represented by the blue dashed line.} }\label{fig:L_f_graphs}
\end{figure}

\subsubsection*{Misspecified Model Example}\label{parametric_example2}
Once again, we now consider fitting our  piecewise linear model to epidemics generated by a parametric model with a power-law kernel, across a population of size N = 250 located at positions $(x_1,y_1)\ldots(x_{250},y_{250})$, generated from independent uniform distributions $ x \sim U[0,50]$ and $y \sim U[0,50]$. The specific epidemic generating model is given by equation \ref{powerlawILM}. For illustration, we show results for one typical epidemic (Figure \ref{epi_curve}, see Appendix) of those generated from this model using the parameter values $\alpha = 0.30$ and $\beta = 2$. We fit a variety of piecewise linear kernel ILMs to this data. We consider one, two, and three change point models with different smoothing prior scale parameters for piecewise linear kernel model with fixed $\delta$ values. Different values of $D$ were chosen to illustrate the effect of the smoothing prior. We used ``vague" positive half-normal  priors for the $\alpha$'s ($\alpha_l \sim N^{+}(0,10^5)$; $l = 1,2,\ldots, n$) and $\beta$'s ($\beta_l \sim N^{-}(0,10^5)$; $l = 1,2,\ldots, n$) for fixed change points. Epidemics are simulated for $T= 49$ time units. The posterior mean, median and $95\%$ PIs for the model parameters and the posterior predictive distribution of the epidemic curve are estimated. We also consider choosing between fitted models with differing change points and/or number of change points using the DIC. The estimated posterior median and $95\%$ PIs were $\alpha = 0.32 ~(0.24,0.43)$ and $\beta = 2.01 ~(1.85,2.18)$ with DIC $1257$, when the true parametric model parameters fitted to the data set generated from the same kernel.\par

Figure \ref{fig:linear_smoothing} (see Appendix) shows the true kernel and fitted piecewise linear kernel based models for different fixed change points with two different smoothing priors scale parameters under the posterior median, and $95\%$ PIs. Figure \ref{fig:linear_smoothing}(a) shows that the true power-law kernel's parameters estimate is very good as we can expect. We can also see that the change point models fitted (Figures \ref{fig:linear_smoothing}((b)-(i)) offer as a reasonably good approximation to the true model under the posterior median. Moreover, we can see that the linear pieces of the kernel are more closely connected to each other at the change point with smaller $95\%$ PIs under the smoothing prior with a lower scale parameter compared to that for the smoothing prior with a larger scale parameter. In addition, the fit of the estimated kernel tends to improve as the number of change points increases. It would appear that the piecewise linear kernel with suitable smoothing prior scale parameter and with appropriate number of fixed change points can approximate the power-law kernel reasonably well, though with higher posterior uncertainty.\par

Table \ref{tab:parametric_linear} shows that the DIC values for the piecewise linear kernel model with different change points and smoothing prior scale parameter values. We see that the DIC values for the ILMs with two change points ($\delta_1 = 2$, and $\delta_2 = 4$) are smaller than the other piecewise linear kernel models suggesting that the single change point models are too inflexible to capture the spatial dynamics of the epidemic and that the models with three change points are unnecessary flexible.\par
The posterior predictive distribution of the epidemic curve form one typical epidemic simulation under true parametric model and other piecewise linear kernel models with fixed change points with two different smoothing scale parameters are shown in Figure \ref{fig:L_fixed_PI} (see Appendix). We can see that the true epidemic curve is very well enveloped by the $95\%$ PIs of the epidemic curve under both the power-law kernel model, and the various piecewise linear kernel models with fixed change points. This suggests that epidemic simulation is reasonably robust to the exact choice of piecewise linear kernel, and that our approach is flexible enough to capture epidemic dynamics when the true kernel is based on a spatial power-law.\par

\begin{table}[h]
     \centering
\small
\begin{tabular}{ | >{\centering\arraybackslash}m{3.0cm} |>{\centering\arraybackslash}m{2.0cm} |>{\centering\arraybackslash}m{3.0cm} |>
{\centering\arraybackslash}m{2.0cm} |}
  
  \hline
  Model &   Change point  $\delta$ & Smoothing parameter & DIC \\
  \hline
  Parametric & -- & -- & 1257 \\
  \hline
  One change point & $2.5$ & $0.037$ &  $1298$ \\
  \hline
  One change point & $2.5$ & $0.003$ &  $1298$ \\
  \hline
  One change point & $ 4$ & $0.037$ & $1298$ \\
  \hline
  One change point & $ 4$ & $0.003$ & $1304$ \\
  \hline
  Two change points & $ (2,4)$ & $(0.047,0.037)$ & $1285$ \\
  \hline
  Two change points & $(2,4)$ & $(0.01,0.01)$ & $1288$ \\
 \hline
  Three change points & $ (2.5,3.5,5)$ & $(0.05,0.05,0.05)$ & $1293$ \\
  \hline
  Three change points & $ (2.5,3.5,5)$ & $(0.01,0.01,0.01)$ & $1293$ \\
  \hline
  \end{tabular}
     \caption{DIC values for power-law model and different semi-parametric ILMs with piecewise linear kernel when the change point ($\delta$) is fixed.}
     \label{tab:parametric_linear}

\end{table}

\section{UK 2001 Foot-and-Mouth Disease (FMD) Data}\label{FMDresults}
Here, we apply the piecewise constant and piecewise linear kernel models described in Section \ref{sec:model2} in an SEIR framework to  foot-and-mouth disease (FMD) data from the $2001$ epidemic in the UK. We compare these results to those under a spatial power-law ILM, which is the model that has more often been considered in the literature. We consider a subset of the UK FMD data from the county of Cumbria in the north-west of England. This subset consists of $1079$ individual cattle and/or sheep farms along with the geographical location in the form of x/y Cartesian coordinates. The farms are considered as the individual-level unit and we ignore the effect of farm size and type.  In the SEIR framework, when a previously susceptible individual farm contracts the infection, they are considered exposed,  or in a  latent period, for some time before they can spread the infection to others.\par 

We consider a fixed and known latent period of $\mu_E = 5$ days, after which the transition from E to I occurs, and an infectious period defined by when the animals on a farm were culled (cull date), which we treat as the removal time. If no cull date is recorded we use an infectious period of $\mu_I = 4$ for the farm.  We consider a temporal subset of the full data consisting of infections for which the recorded exposure times lie between  $t_{min} = 28$ to $t_{max} = 66$ days after the first infection, which occurred on February 19, 2001.

The form of the model in equation (\ref{meth.sec.2.3.eq4}) used is slightly modified to include a constant so-called sparks term, $\epsilon(i,t) = \varepsilon.$ Specifically, our model is: 

\begin{align}\label{fmd.model}
    P_{it}(\theta) = 1 - \exp{\left\{-\left(\sum_{j \in I_t}\tilde{k}(i,j)+ \varepsilon \right)\right\}}
\end{align}

\noindent The purpose of the sparks term is to allow for a small, random probability of a farm being infected at any given time. This allows for the fact that we are observing a subset of the UK foot-and-mouth disease data, and so infections can arise, seemingly at random, from outside the observed population.

Under our SEIR framework, the form of the likelihood is given by equations (\ref{meth.sec.2.2.eq2}) and (\ref{meth.sec.2.2.eq3}). For piecewise constant kernel models, we used ``vague" positive half-normal priors for the $\alpha$'s ($\alpha_l \sim N^{+}(0,10^5)$; $l = 1,2,\ldots, n$) for fixed and unknown change points. When the change points are unknown, for the two-step case, we use ``vague" prior $\delta_1 \sim U(0,7)$, and for the two three-step cases we use ``weakly" informative priors $\delta_1 \sim U(1,3)$, $\delta_2 \sim U(3,5)$ and $\delta_1 \sim U(1,5)$, $\delta_2 \sim U(5,8)$, respectively. Here, weakly informative priors are chosen due to the potential non-identifiability associated with using vague priors. Again, for piecewise linear kernel models, we used ``vague" positive half-normal priors for the $\alpha$'s ($\alpha_l \sim N^{+}(0,10^5)$; $l = 1,2,\ldots, n$) and $\beta$'s (e.g., $\beta_l \sim N^{-}(0,10^5)$; $l = 1,2,\ldots, n$) for fixed change points, along with the smoothing prior. Here again, two different values of scale parameter $D$ were chosen to illustrate the effect of the smoothing prior on the one change point linear kernel model. We evaluate MCMC convergence using traceplot visual inspection and Geweke's diagnostic (\cite{geweke1992evaluating}) for each model parameter. If any parameters fail to converge based on Geweke's test, we employ the Gelman-Rubin diagnostic (\cite{gelman1992inference}) across three MCMC chains.


\subsection{Results}
\begin{figure}[h!tpb]
      \centering
	   \begin{subfigure}{0.24\linewidth}
		\includegraphics[width=\linewidth]{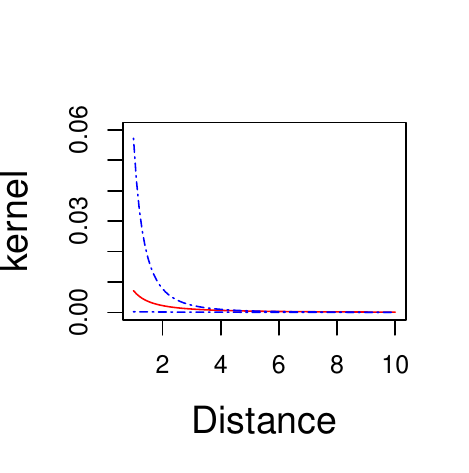}
		\caption{}
		\label{fig:subfig1}
	   \end{subfigure}
	   \begin{subfigure}{0.24\linewidth}
		\includegraphics[width=\linewidth]{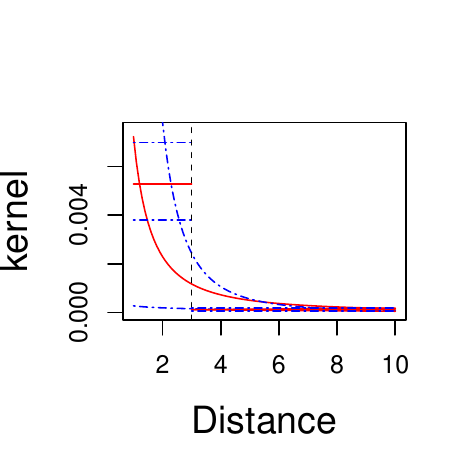}
		\caption{$\delta = 3$}
		\label{fig:subfig2}
	    \end{subfigure}
        \begin{subfigure}{0.24\linewidth}
		\includegraphics[width=\linewidth]{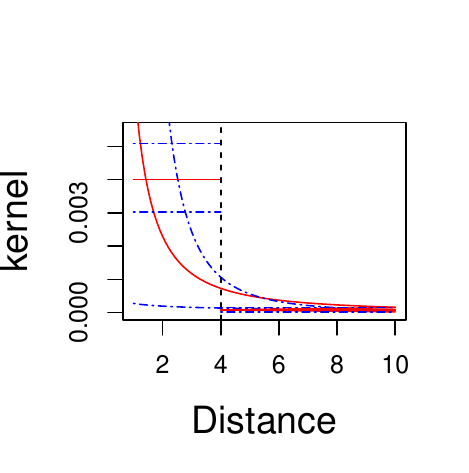}
		\caption{$\delta = 4$}
		\label{fig:subfig2}
	    \end{subfigure}
	\vfill
	     \begin{subfigure}{0.24\linewidth}
		 \includegraphics[width=\linewidth]{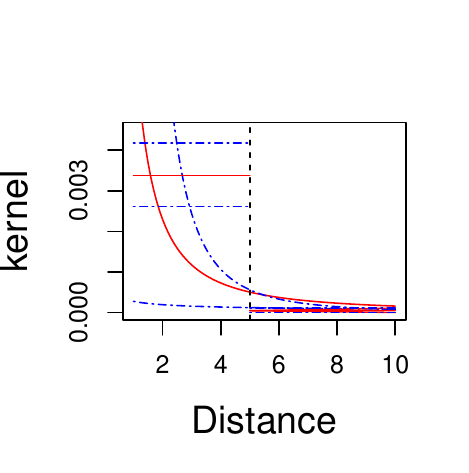}
		 \caption{$\delta = 5$}
		 \label{fig:subfig3}
	      \end{subfigure}
	       \begin{subfigure}{0.24\linewidth}
		  \includegraphics[width=\linewidth]{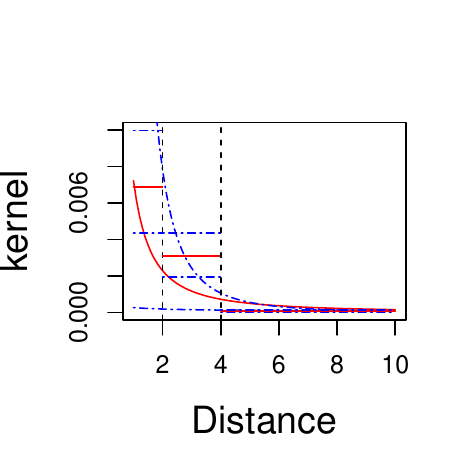}
		  \caption{$\delta = (2,4)$}
		  \label{fig:subfig4}
	       \end{subfigure}
        \begin{subfigure}{0.24\linewidth}
		  \includegraphics[width=\linewidth]{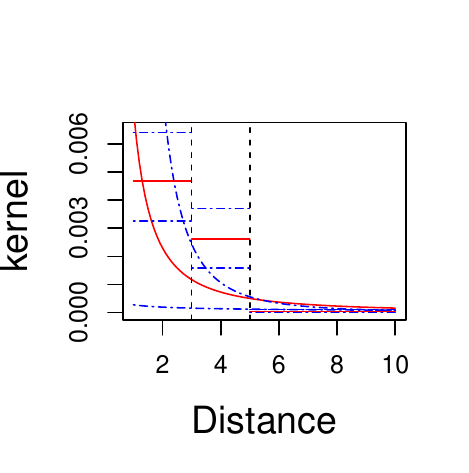}
		  \caption{$\delta = (3,5)$}
		  \label{fig:subfig4}
	       \end{subfigure}
        \vfill
        \begin{subfigure}{0.24\linewidth}
		  \includegraphics[width=\linewidth]{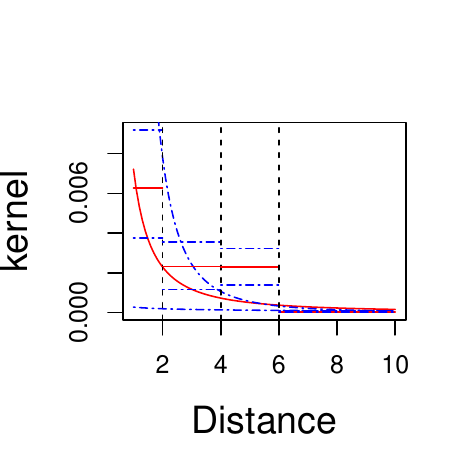}
		  \caption{$\delta = (2,4,6)$}
		  \label{fig:subfig4}
	       \end{subfigure}
        \begin{subfigure}{0.24\linewidth}
		  \includegraphics[width=\linewidth]{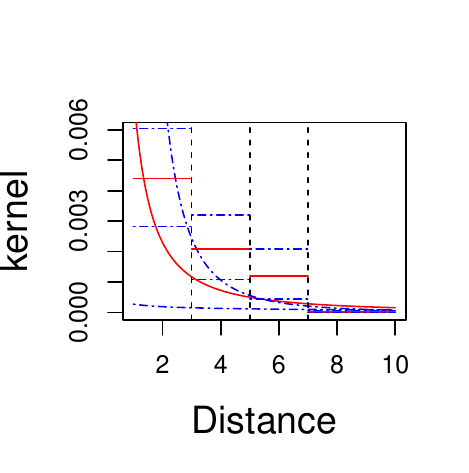}
		  \caption{$\delta = (3,5,7)$}
		  \label{fig:subfig4}
	       \end{subfigure}
        \begin{subfigure}{0.24\linewidth}
		  \includegraphics[width=\linewidth]{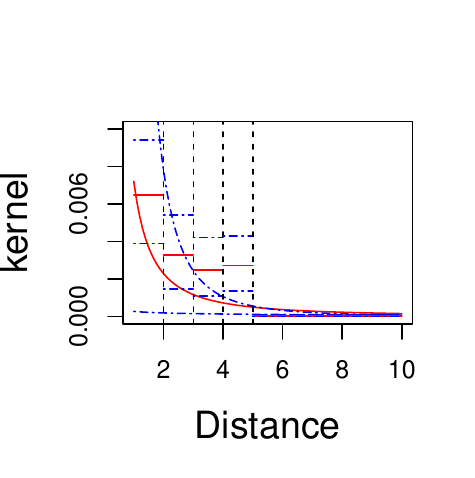}
		  \caption{$\delta = (2,3,4,5)$}
		  \label{fig:subfig4}
	       \end{subfigure}
        \vfill
        \begin{subfigure}{0.24\linewidth}
		  \includegraphics[width=\linewidth]{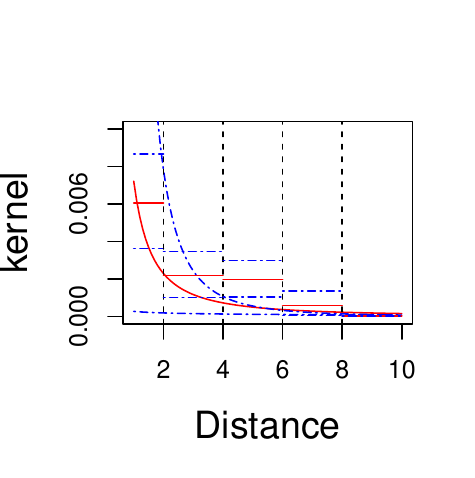}
		  \caption{$\delta = (2,4,6,8)$}
		  \label{fig:subfig4}
	       \end{subfigure}
	\caption{\small{Fitted kernel under the posterior median and $95\%$ PIs for (a) the power-law kernel model are represented by red (solid), and blue dashed lines, respectively. Posterior medians and $95\%$ PIs for different change point ($\delta$) models (b-j) with piecewise constant kernel which fitted to FMD data are represented by red  and blue dashed lines, respectively.}}\label{fig:FMD_para_fix}
\end{figure}

Figure \ref{fig:FMD_para_fix} shows the fitted power-law kernel and  piecewise constant kernel models for different fixed change points under the posterior, when the models have been fitted to the 2001 UK FMD data. The estimated posterior median (and $95\%$ PIs) for the power-law kernel model parameters were: $\alpha_{PM} = 0.007 ~   (0.00028,0.057), \beta_{PM} = 1.65 (0.51,2.87)$ and $\varepsilon_{PM} = 0.0073 ~ (0.00035,0.031),$ with a DIC of $3889$.\par
We can see that the fitted piecewise constant kernels mimic the fitted power-law about as well as such a function can. Further, this mimicry of the fitted power-law kernel remains as the number of change points varies.
We can also see that the $95\%$ PI of power-law kernel model  gets narrower as the distance between individual increases, and we see a similar narrowing of the $95\%$ PIs under the piecewise constant kernel models.  

Table \ref{tab:FMDparametric_fixed} shows the DIC values under power-law and  piecewise constant kernels models. We can see that in all cases the piecewise constant kernels models produce a lower DIC suggesting a better fit to the data.
Further, we observe that the DIC values under the piecewise constant kernels ILMs tend become smaller as the number of change points increase, suggesting the greater flexibility of these models is favoured. In fact, the two smallest DIC values $2733$ and $2734$ are observed for the five-step model with change points at distances $2, 4, 6, 8$ and four-step model with change points $2, 4, 6$, respectively.\par

Results for the piecewise constant kernel ILMs with unknown change points are shown in Figure \ref{fig:FMD_para_varying} (see Appendix). 
We see similar results, but with a greater degree of posterior uncertainty in the estimated kernel. The DIC value of the model with two unknown change points of $2731$ is lower than that of any of the models with fixed change points, implying that this extra flexibility may improve the fit (Table \ref{tab:FMD_parametric_varying}). However, this is only a reduction of $2$ or $3$ from the DIC of the two fixed change point models above, suggesting that either approach would be reasonable.\par  

Finally, Figure \ref{fig:FMD_f_linear_smoothing} (see Appendix) shows the fitted power-law and piecewise linear kernel models for one fixed change point model with two different smoothing prior scale parameters under the posterior. We can see that the ends of the linear pieces of the kernel tend to become closer under the smoothing prior with scale parameter $D = 0.10$. Moreover, the DIC values (Table \ref{tab:fmd_f_parametric_linear}) under the piecewise linear kernel with both scale parameters are similar and much smaller than that under the power-law kernel model. This implies that the one change point linear kernel models may give a better fit to the 2001 UK FMD data here.\par

The fact that the DIC tends to suggest that the piecewise kernels with larger number of steps tend to fit better, and certainty better than the simple power-law kernel, is interesting epidemiologically. Of course, the spatial kernel needs to capture the effects of different transmission mechanisms which may occur at different frequencies over different spatial scales. For example, FMD may be transmitted by mechanisms such as different wildlife species carrying the disease (eg., mice, birds, foxes), wind plumes, people or vehicle movements, livestock movements and/or animal contact through a fence. In any case, our results suggest that the simple parametric approach usually taken to modeling spatial transmission risk may be sub-optional.

\begin{table}[h]
     \centering
\small
\begin{tabular}{ | >{\centering\arraybackslash}m{2.0cm} |>{\centering\arraybackslash}m{2.cm} |>{\centering\arraybackslash}m{3.0cm} | }

  \hline
  Model &   Fixed $\delta$ & DIC \\
  \hline
  Parametric & & 3889 \\
  \hline
  Two-step  & 3 &  2775 \\
  \hline
  Two-step & 4 & 2765  \\
  \hline
  Two-step & 5 & 2749  \\
  \hline
  Three-step & (2,4) & 2761  \\
  \hline
  Three-step  & (3,5) & 2747  \\
  \hline
  Four-step & (2,4,6) & 2734  \\
  \hline
  Four-step & (3,5,7) & 2738  \\
  \hline
  Five-step & (2,3,4,5) & 2747  \\
   \hline
   Five-step & (2,4,6,8) & 2733  \\
   \hline

  \end{tabular}
     \caption{DIC values for power-law model and different semi-parametric ILMs with piecewise constant kernel when the change point ($\delta$) is fixed and fitted to FMD data.}
     \label{tab:FMDparametric_fixed}
  \end{table}

\begin{table}[h]
     \centering
\small
\begin{tabular}{ | >{\centering\arraybackslash}m{2.0cm} |>{\centering\arraybackslash}m{5.0cm} |
>{\centering\arraybackslash}m{3.0cm} | }
  
  \hline
  Model &   Estimated  $\delta$ ($95\% PI$) & DIC \\
  \hline
  Parametric & & 3889 \\
  \hline
  Two-step & $\delta_1 = $6.23  (5.47,6.54) &  2737 \\
  \hline
  Three-step & $\delta_1 = $1.90  (1.24,2.56), ~ $\delta_2 = $4.87  (4.47, 4.99) & 2740  \\
  \hline
  Three-step & $\delta_1 = $2.13(1.61,3.47), $\delta_2 = $6.32(5.60,7.62) & 2731  \\
  \hline
  
  \end{tabular}
     \caption{DIC values for power-law model and different semi-parametric ILMs with piecewise constant kernel when the change point ($\delta$) is estimated and fitted to FMD data.}
     \label{tab:FMD_parametric_varying}
 
    \end{table}

\begin{table}[h]
     \centering
\small
\begin{tabular}{ | >{\centering\arraybackslash}m{3.0cm} |>{\centering\arraybackslash}m{2.0cm} |>{\centering\arraybackslash}m{3.0cm} |>
{\centering\arraybackslash}m{2.0cm} |}
  
  \hline
  Model &   Change point  $\delta$ & Smoothing parameter & DIC \\
  \hline
  Parametric & -- & -- & 3889 \\
  \hline
  Two-step  & $3$ & $0.10$ &  2719 \\
  \hline
  Two-step  & $3$ & $0.04$ &  2719 \\
  \hline
  
  \end{tabular}
     \caption{DIC values for power-law model and different semi-parametric ILMs with piecewise linear kernel when the change point ($\delta$) is fixed and fitted to FMD data.}
     \label{tab:fmd_f_parametric_linear}

\end{table}

\section{Discussion}\label{discussion}
In this paper, we develop a framework for basic piecewise spatial infectious disease individual-level models in a Bayesian MCMC framework. 
Under a number of simulated epidemic scenarios, we illustrate that the true parameters of the ILMs with a piecewise constant or piecewise linear kernel, with either for fixed or estimated change points, can be successfully estimated. In addition, we found that the piecewise constant kernel with the appropriate number of fixed or estimated change points with significant size, the $95\%$ PIs, can approximate the kernel of a misspecified model well. The piecewise linear kernel ILM with a suitable smoothing prior showed  similar results. Moreover, these models have good posterior predictive ability and exhibit the capability of capturing the true epidemic curve, even when the fitted model is  misspecified to some degree. However,   we did observe that carrying out MCMC becomes increasingly  burdensome as the number of model parameters increased, especially when the change points were estimated. \par

Future research could take us in numerous different directions. We have examined here relatively simple non-parametric functions for use as spatial transmission kernels. However, it would also be possible to other forms of spline (eg., cubic splines or B-splines (\cite{silverman1985some,friedman2001elements})) or Gaussian processes. The latter approach was successfully implemented by \cite{kypraios2018bayesian}, for example. A systematic comparison of the performance of different non-parametric spatial kernel would also be warranted.\par 

We can also explore other methods to estimate the number of change points. Our work here suggests that the highly popular DIC might be a good option, but other criterion could be considered. We might also want to use our posterior predictive approach to develop posterior predictive p-values, likely for scalar metrics such as the epidemic peak time or intensity (e.g., \cite{gardner2011goodness}). We could also consider including the number of change points as a parameter to be estimated and use reversible jump MCMC (e.g., \cite{green2009reversible}). Furthermore, work needs to be done on choosing the smoothing parameter in the case of the piecewise linear kernel ILMs. Here, we took an ad-hoc approach, simply observing the effect of changing the smoothing parameter ($D$). Perhaps an obvious place to start would be to build up a hierarchical structure, putting a prior on the smoothing parameter, allowing the data to inform it. Moreover, we could consider more complex forms for the kernel, such as  piecewise quadratic or cubic kernels to estimate the non-linear relationship between spatial distance and the risk of infection more accurately.\par

In our study, we considered models with no covariates. In real life, the individual or group-level factors likely have an impact on infection transmission or susceptibility.  We could extend our model by incorporating covariates to consider more complex dynamics. For example, in our foot-and-mouth disease models we ignored the number of cattle and sheep on the farm, since we want to focus on the spatial aspect of disease spread. However, these are often used as covariates in such models (\cite{deardon2010inference,ward2023framework}).\par

It would also be of interest to develop semi-parametric forms of other epidemic models. For example, we could also consider including a so-called alarm function in our model to allow for the effect of disease prevalence-related behavioural change in the population over time (\cite{ward2023framework, ward2023bayesian}) .
Alternatively, the so-called geographically-dependent ILMs of \cite{mahsin2022geographically}, which allow for both an individual and regional level spatial structure in the model, could be extended to allow for our piecewise spatial kernels.\par

Moreover, we considered models in which we had fixed and known latent and/or infectious periods to contain computational burden and to keep our analysis simple.  
However, in many scenarios this is an unrealistic assumption, since such periods can vary greatly between individuals, and often the times at which individuals are exposed, become infectious and/or are removed will be censored. Such uncertainty can be allowed for using techniques such as data augmented MCMC, although this would substantially increase the computational burden. In addition, a limiting assumption in our spatial ILMs is that we assume we know the spatial locations of individuals with certainty. This is obviously not true for humans, since people move around, but also questionable for the FMD example. The dataset contains map coordinates for farmhouses which we used as proxy for the location of animals. Of course, in reality animals may actually be located quite far from a farmhouse. \cite{deardon2012spatial} introduced parametric spatial measurement error models to try and account for this uncertainty, and it would be of interest to extend this approach to our piecewise, and other semi-parametric, ILMs.\par

Finally, given that Bayesian analysis of ILMs are certainly computationally burdensome for large populations, it would be of utility to explore the various approximate methods of carrying out inference available to reduce such burdens. For example, \cite{kwong2012linearized} have already shown that when event times are known we can use the linear nature of kernels such as the piecewise linear kernel, to lower computation times by separating part of the likelihood into an easy to compute parameter dependent component, and a much more intensive but parameter independent component which only needs to be computed once. Unfortunately, this does not work if we are using techniques such as data augmentation, but we might consider approximate Bayesian computation methods (e.g., in the context of ILMs, \cite{almutiry2019incorporating,ward2023bayesian}),  or Gaussian process emulation (e.g., \cite{pokharel2016gaussian,pokharel2022emulation}).
\newpage
\section*{Appendix}\label{appendix}
\begin{figure}[H]
    \centering
    \begin{subfigure}{0.32\linewidth}
    \includegraphics[width=\linewidth,trim={0.05cm -0.50cm 0.10cm 0.2cm},clip]{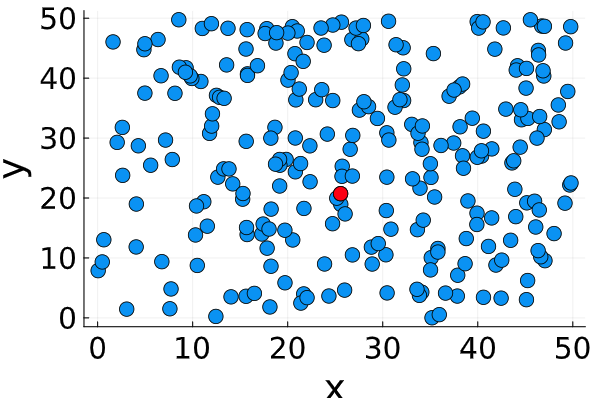}\caption{$t = 1$}
    \end{subfigure}
    \begin{subfigure}{0.32\linewidth}
    \includegraphics[width=\linewidth]{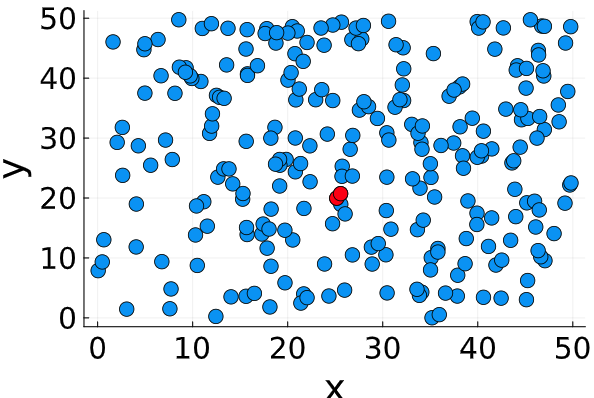} \caption{$t = 2$}
    \end{subfigure}
    \begin{subfigure}{0.32\linewidth}
        \includegraphics[width=\linewidth,trim={0.05cm -0.20cm 0.10cm 0.2cm},clip]{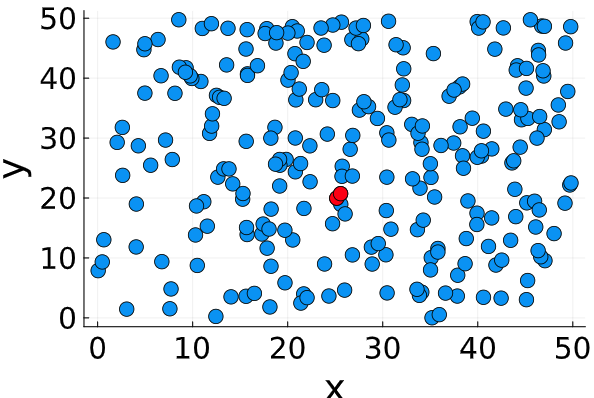}\caption{$t = 3$}
     \end{subfigure}
     \begin{subfigure}{0.32\linewidth}
        \includegraphics[width=\linewidth,trim={0.05cm -0.20cm 0.10cm 0.2cm},clip]{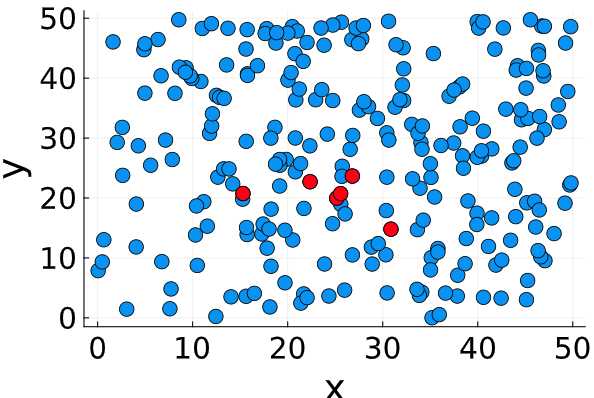}\caption{$t = 4$}
     \end{subfigure}
     \begin{subfigure}{0.32\linewidth}

        \includegraphics[width=\linewidth,trim={0.05cm -0.20cm 0.10cm 0.2cm},clip]{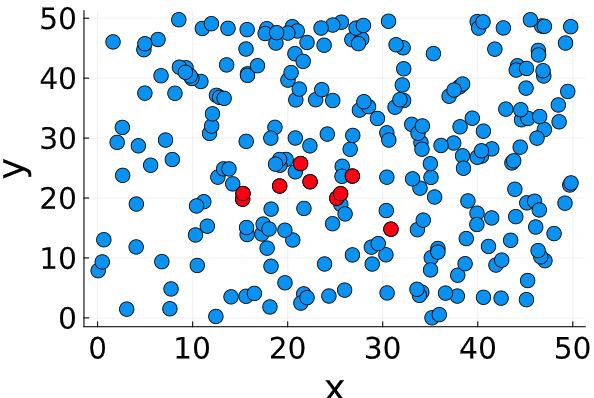}\caption{$t = 5$}
     \end{subfigure}
      \begin{subfigure}{0.32\linewidth}

        \includegraphics[width=\linewidth,trim={0.05cm -0.20cm 0.10cm 0.2cm},clip]{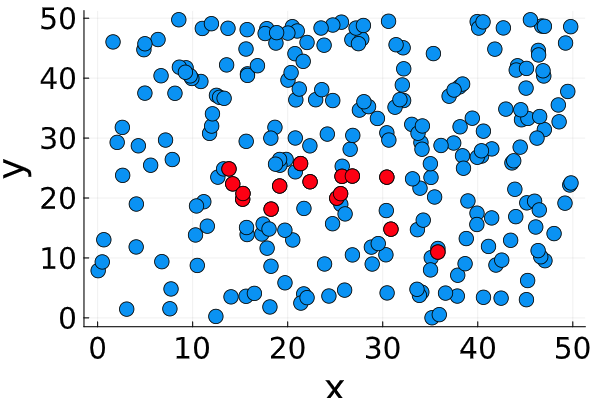}\caption{$t = 6$}
     \end{subfigure}
     \begin{subfigure}{0.32\linewidth}
        \includegraphics[width=\linewidth,trim={0.05cm -0.20cm 0.10cm 0.2cm},clip]{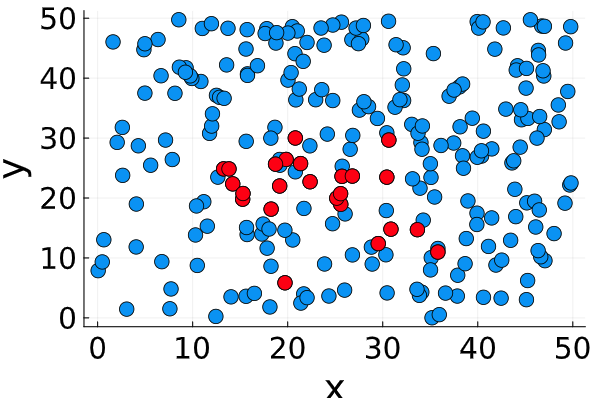}\caption{$t = 7$}
     \end{subfigure}
      \begin{subfigure}{0.32\linewidth}

        \includegraphics[width=\linewidth,trim={0.05cm -0.20cm 0.10cm 0.2cm},clip]{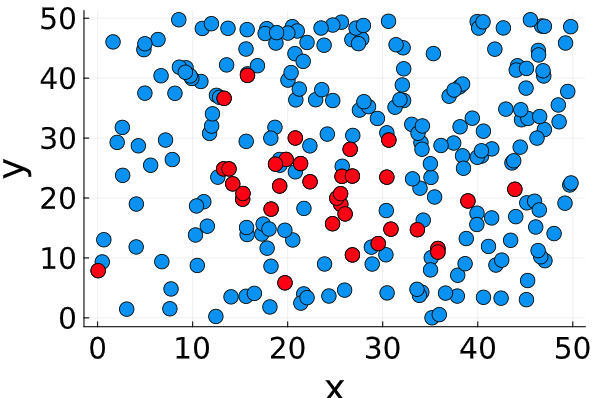}\caption{$t = 8$}
     \end{subfigure}
     \begin{subfigure}{0.32\linewidth}
        \includegraphics[width=\linewidth,trim={0.05cm -0.20cm 0.10cm 0.2cm},clip]{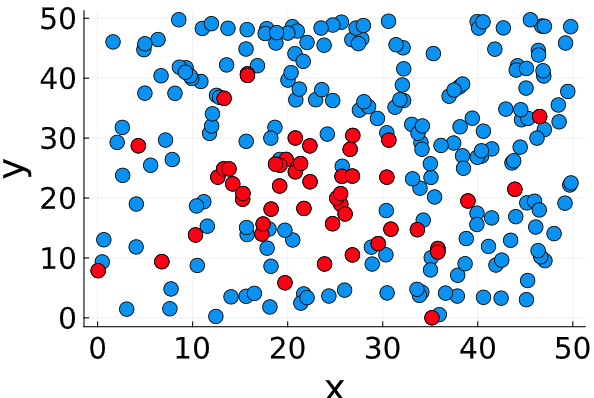}\caption{$t = 9$}
     \end{subfigure}
     \begin{subfigure}{0.32\linewidth}
        \includegraphics[width=\linewidth,trim={0.05cm -0.20cm 0.10cm 0.2cm},clip]{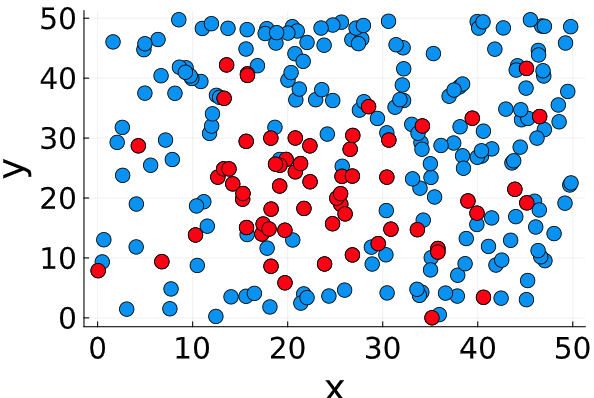}\caption{$t = 10$}
     \end{subfigure}
      \begin{subfigure}{0.32\linewidth}

        \includegraphics[width=\linewidth,trim={0.05cm -0.20cm 0.10cm 0.2cm},clip]{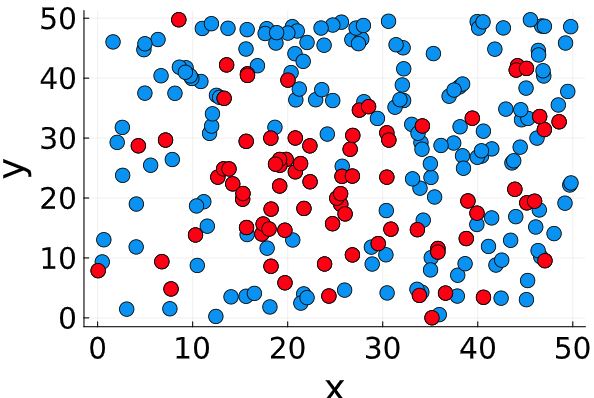}\caption{$t = 11$}
     \end{subfigure}
      \begin{subfigure}{0.32\linewidth}

        \includegraphics[width=\linewidth,trim={0.05cm -0.20cm 0.10cm 0.2cm},clip]{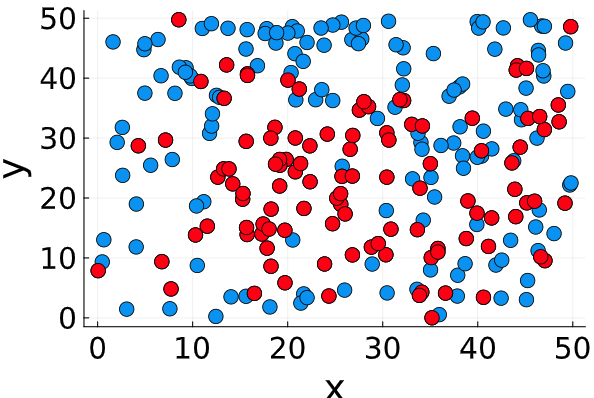}\caption{$t = 12$}
     \end{subfigure}
     \begin{subfigure}{0.32\linewidth}
        \includegraphics[width=\linewidth,trim={0.05cm -0.20cm 0.10cm 0.2cm},clip]{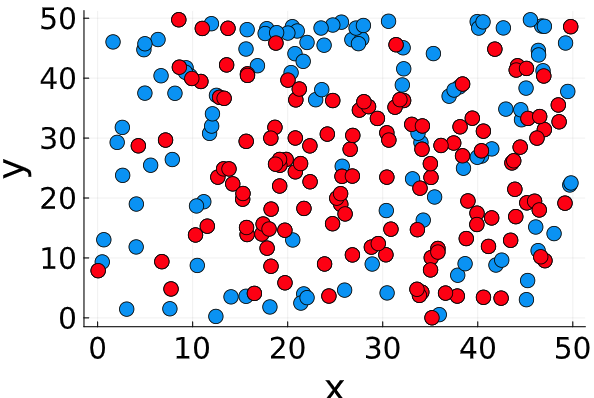}\caption{$t = 13$}
     \end{subfigure}  
     \begin{subfigure}{0.32\linewidth}
        \includegraphics[width=\linewidth,trim={0.05cm -0.20cm 0.10cm 0.2cm},clip]{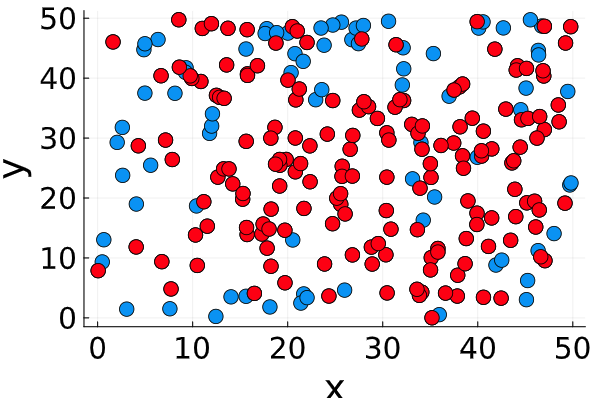}\caption{$t = 14$}
     \end{subfigure}
     \begin{subfigure}{0.32\linewidth}
        \includegraphics[width=\linewidth,trim={0.05cm -0.20cm 0.10cm 0.2cm},clip]{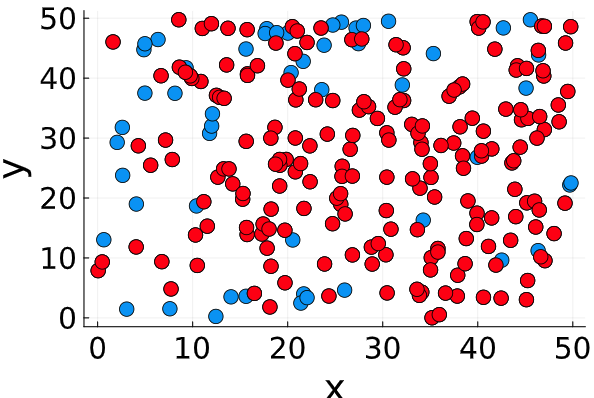}\caption{$t = 15$}
     \end{subfigure}
     \caption{\small{Typical simulated epidemic realization across a grid of individuals, where susceptible individuals are denoted by blue circles and infected individuals are denoted by red circles.} }\label{epi_curve}
\end{figure}

\begin{figure}[h!tpb]
    \includegraphics[width=\textwidth]{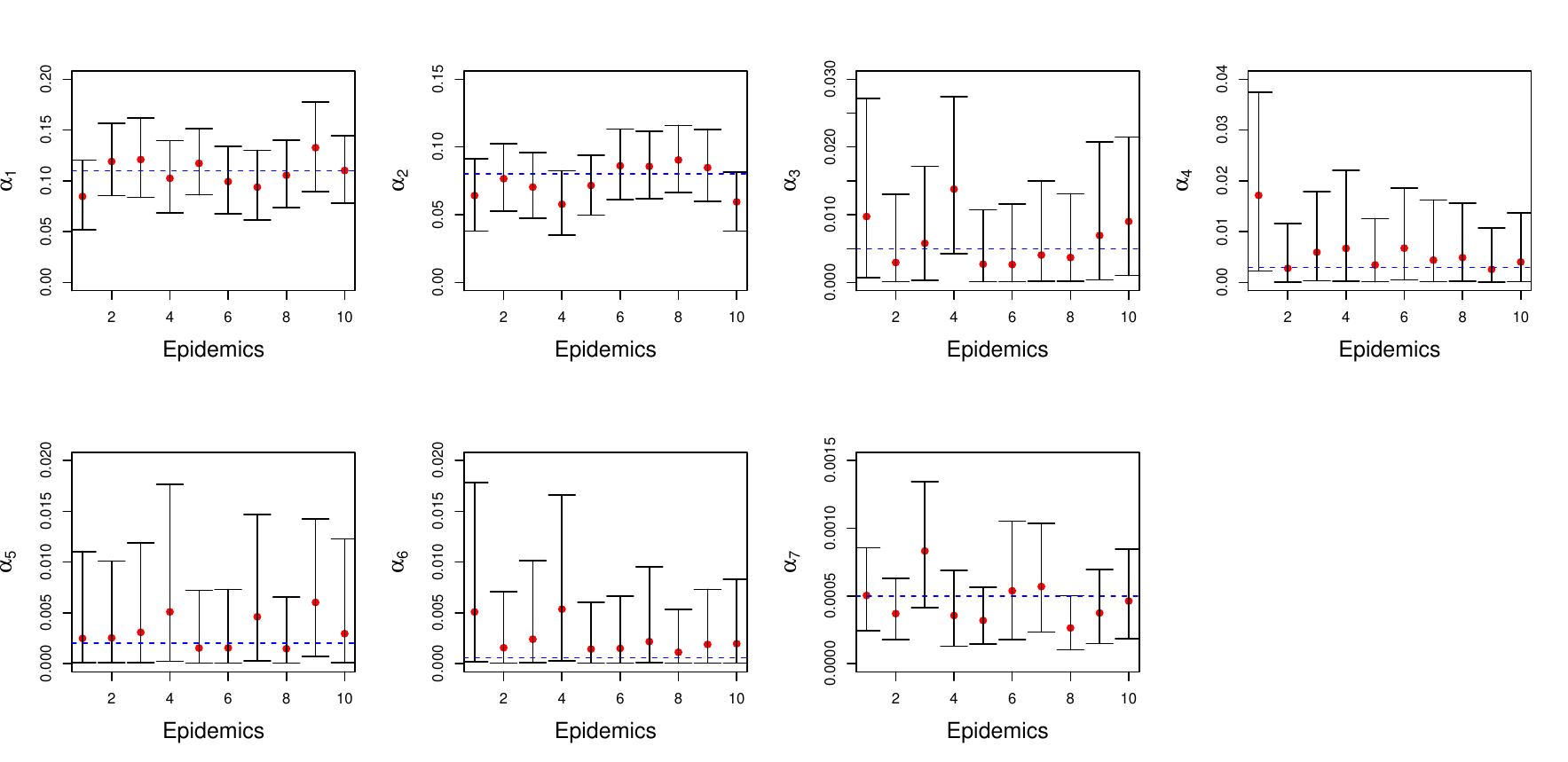}
     \caption{\small{Posterior medians (red points) and $95\%$ PIs for  $\alpha_1$ to $\alpha_7$ for $10$ different simulated epidemics when the change points ($\delta$'s) are considered fixed. The true parameter values $\alpha_1 = 0.11 ,\alpha_2 = 0.08, \alpha_3 = 0.005, \alpha_4 = 0.003,\alpha_5 = 0.002,\alpha_6 = 0.0006$, and $\alpha_7 = 0.0005$ for seven-step cases with fixed $\delta_1 = 2.5, \delta_2 = 4, \delta_3 = 5.5, \delta_4 = 6.5,\delta_5 = 7.5,\delta_6 = 8.5$ and $\delta_7 = 9.5$ are represented by the blue dashed line.} }\label{fig:7_f_graphs}
\end{figure}

\begin{figure}[h!tpb]
    \centering
    \includegraphics[width=\textwidth]{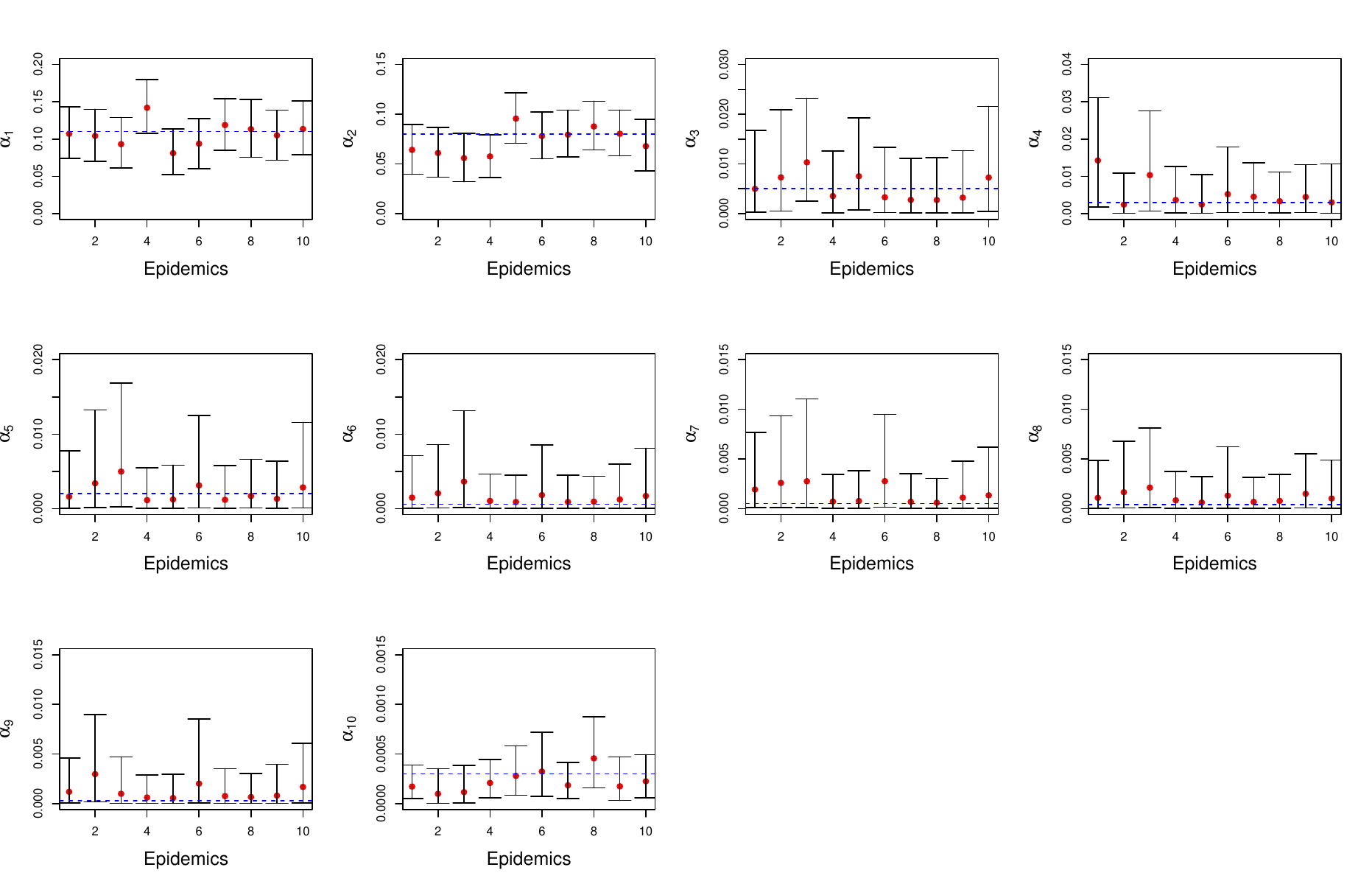}
     \caption{\small{Posterior medians (red points) and $95\%$ PIs for $\alpha_1$ to $\alpha_{10}$ for $10$ different simulated epidemics when the change points ($\delta$'s) are considered fixed. The true parameter values $\alpha_1 = 0.11 ,\alpha_2 = 0.08, \alpha_3 = 0.05, \alpha_4 = 0.003,\alpha_5 = 0.002,\alpha_6 = 0.0006, \alpha_7 = 0.0005,\alpha_8 = 0.0004,\alpha_9 = 0.0003$ and $\alpha_{10} = 0.0003$ for ten-step cases with fixed $\delta_1 = 2.5, \delta_2 = 4, \delta_3 = 5.5, \delta_4 = 6.5,\delta_5 = 7.5,\delta_6 = 8.5$ and $\delta_7 = 9.5, \delta_8 = 10.5$, and $\delta_9 = 11.5$ are represented by the blue dashed line.} }\label{fig:10_f_graphs}
\end{figure}

\begin{figure}[h!tpb]
      \centering
	   \begin{subfigure}{0.325\linewidth}
		\includegraphics[width=\linewidth]{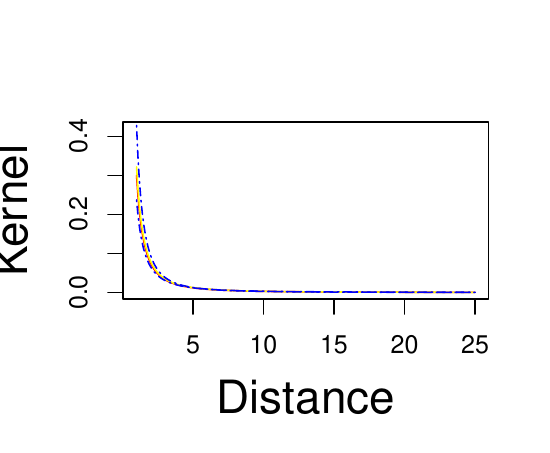}
		\caption{}
		\label{fig:subfig1}
	   \end{subfigure}
	   \begin{subfigure}{0.325\linewidth}
		\includegraphics[width=\linewidth]{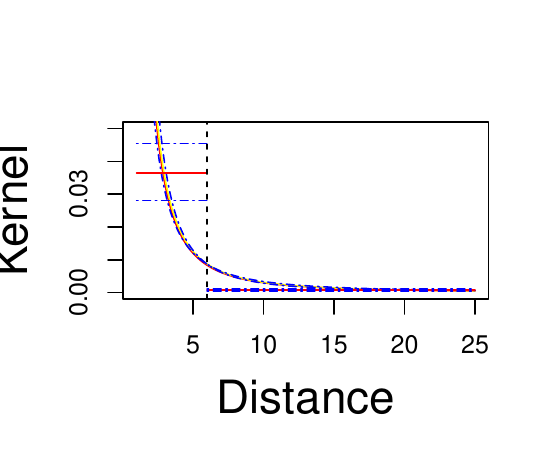}
		\caption{$\delta = 6$}
		\label{fig:subfig2}
	    \end{subfigure}
        \begin{subfigure}{0.325\linewidth}
		\includegraphics[width=\linewidth]{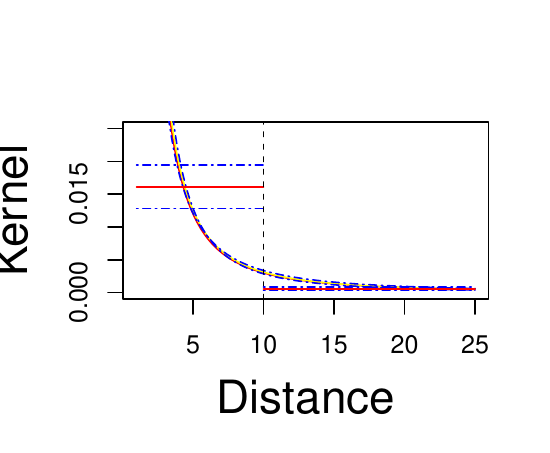}
		\caption{$\delta = 10$}
		\label{fig:subfig2}
	    \end{subfigure}
	\vfill
	     \begin{subfigure}{0.325\linewidth}
		 \includegraphics[width=\linewidth]{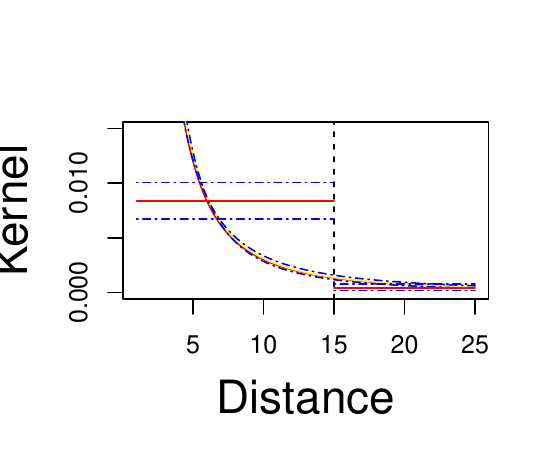}
		 \caption{$\delta = 15$}
		 \label{fig:subfig3}
	      \end{subfigure}
	       \begin{subfigure}{0.325\linewidth}
		  \includegraphics[width=\linewidth]{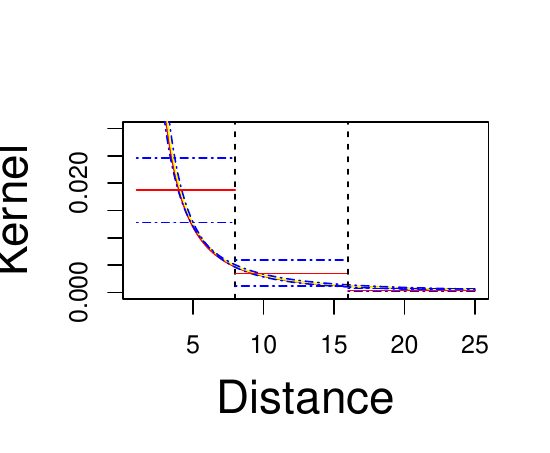}
		  \caption{$\delta = (8,16)$}
		  \label{fig:subfig4}
	       \end{subfigure}
        \begin{subfigure}{0.325\linewidth}
		  \includegraphics[width=\linewidth]{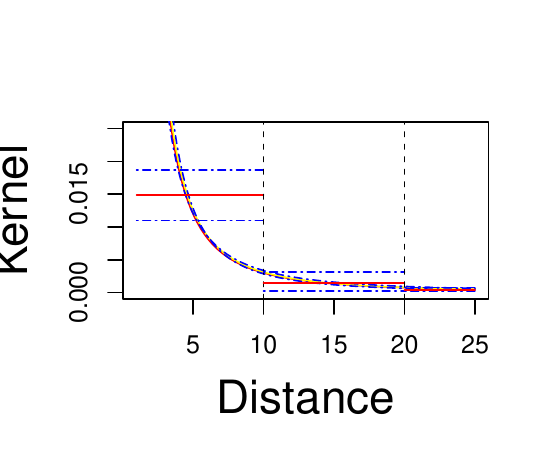}
		  \caption{$\delta = (10,20)$}
		  \label{fig:subfig4}
	       \end{subfigure}
        \vfill
        \begin{subfigure}{0.325\linewidth}
		  \includegraphics[width=\linewidth]{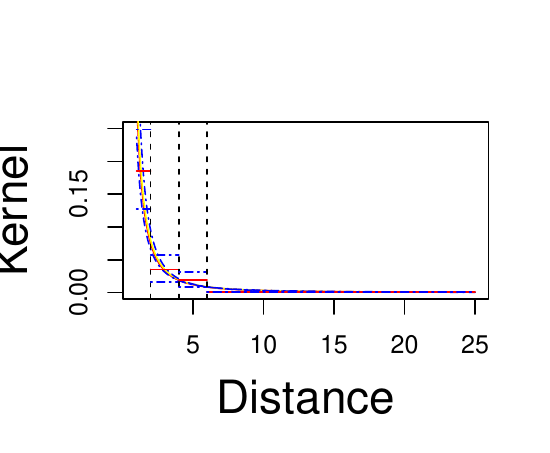}
		  \caption{$\delta = (2,4,6)$}
		  \label{fig:subfig4}
	       \end{subfigure}
        \begin{subfigure}{0.325\linewidth}
		  \includegraphics[width=\linewidth]{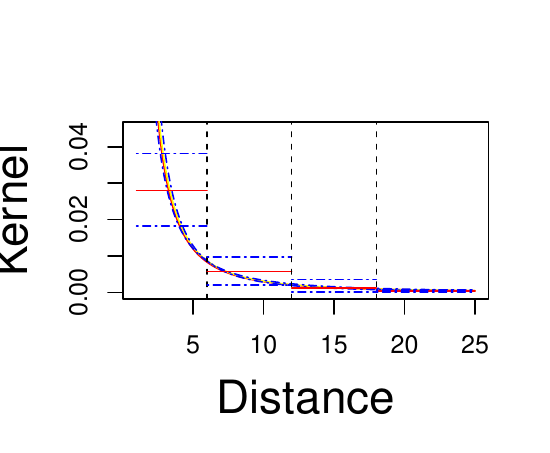}
		  \caption{$\delta = (6,12,18)$}
		  \label{fig:subfig4}
	       \end{subfigure}
        \begin{subfigure}{0.325\linewidth}
		  \includegraphics[width=\linewidth]{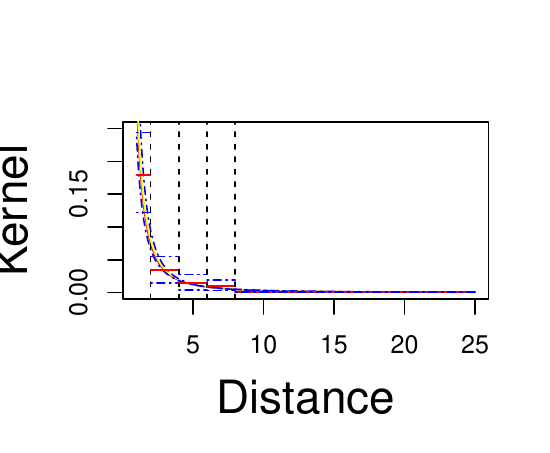}
		  \caption{$\delta = (2,4,6,8)$}
		  \label{fig:subfig4}
	       \end{subfigure}
        \vfill
        \begin{subfigure}{0.325\linewidth}
		  \includegraphics[width=\linewidth]{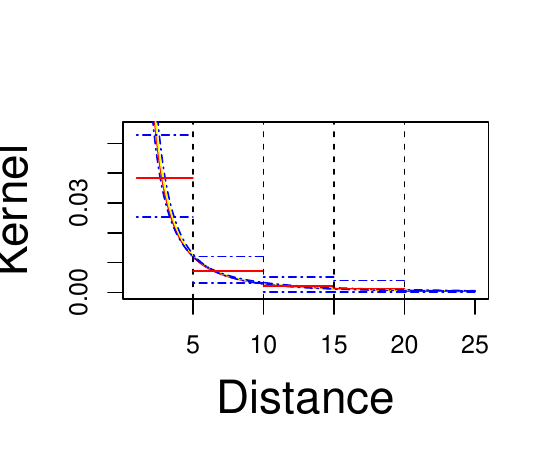}
		  \caption{$\delta = (5,10,15,20)$}
		  \label{fig:subfig4}
	       \end{subfigure}
	\caption{\small{True kernel,  fitted kernel under the posterior median and $95\%$ PIs for (a) the true power-law  model are represented by red (solid), yellow (solid) and blue (dashed) lines, respectively. Posterior medians and $95\%$ PIs for different change point ($\delta$) models (b-j) are represented by red (solid)  and blue (dashed) lines, respectively.}}\label{fig:para_fix}
\end{figure}

\begin{figure}[h!tpb]
      \centering
	   \begin{subfigure}{0.325\linewidth}
		\includegraphics[width=\linewidth]{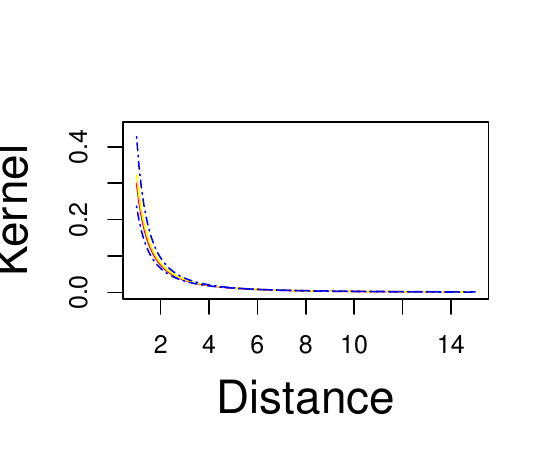}
		\caption{}
		\label{fig:subfig1}
	   \end{subfigure}
	   \begin{subfigure}{0.325\linewidth}
		\includegraphics[width=\linewidth]{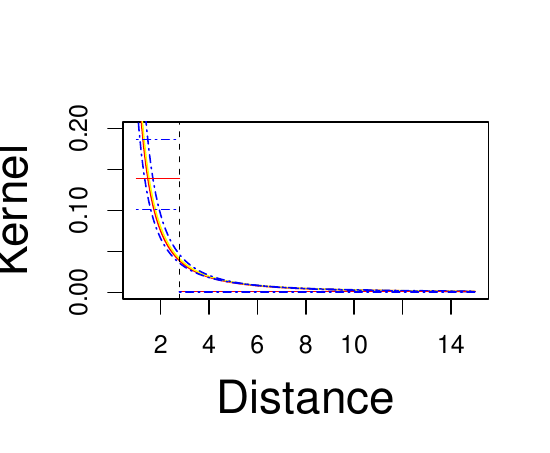}
		\caption{$\delta = 2.78$}
		\label{fig:subfig2}
	    \end{subfigure}
        \begin{subfigure}{0.325\linewidth}
		\includegraphics[width=\linewidth]{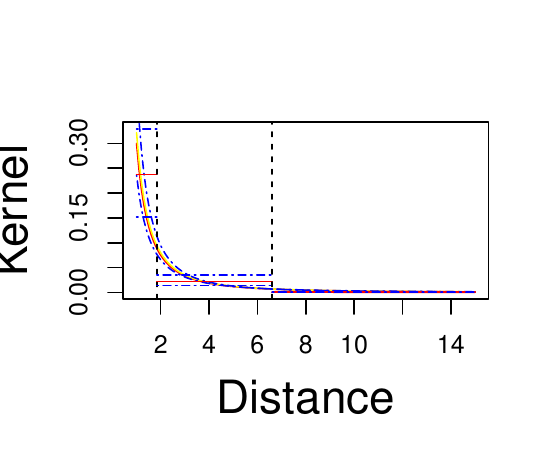}
		\caption{$\delta = (1.83,6.61)$}
		\label{fig:subfig2}
	    \end{subfigure}
     \caption{\small{True kernel,  and fitted kernel under the posterior median and $95\%$ PIs for (a) the true power-law model are represented by red (solid), yellow (solid) and blue (dashed) lines, respectively. Posterior medians and $95\%$ PIs of ILMs (b-c) with estimated posterior median of the change points ($\delta$'s) are represented by red (solid) and blue (dashed) lines, respectively.} }\label{fig:para_varying}
\end{figure}
     

\begin{figure}[h!tpb]
     \begin{subfigure}[b]{0.32\textwidth}
         \centering
         \includegraphics[width=\textwidth]{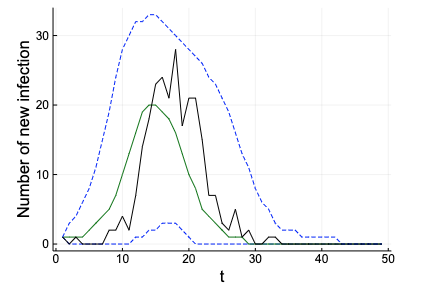}
         \caption{Parametric model}
         \label{fig:y equals x}
     \end{subfigure}
     \hfill
     \begin{subfigure}[b]{0.32\textwidth}
         \centering
         \includegraphics[width=\textwidth]{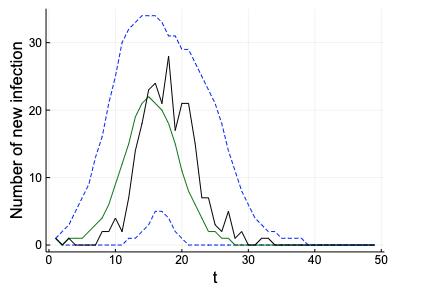}
         \caption{$\delta$ = 6}
         \label{fig:three sin x}
     \end{subfigure}
     \hfill
     \begin{subfigure}[b]{0.32\textwidth}
         \centering
         \includegraphics[width=\textwidth]{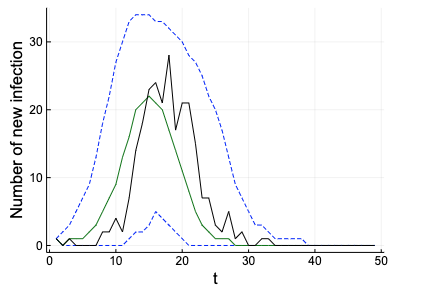}
         \caption{$\delta$ = 10}
         \label{fig:three sin x}
     \end{subfigure}
     \hfill
     \begin{subfigure}[b]{0.32\textwidth}
         \centering
         \includegraphics[width=\textwidth]{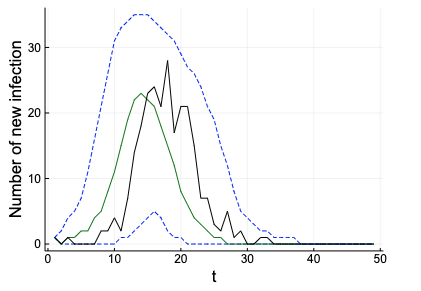}
         \caption{$\delta$ = 15}
         \label{fig:three sin x}
     \end{subfigure}
    \hfill
     \begin{subfigure}[b]{0.32\textwidth}
         \centering
         \includegraphics[width=\textwidth]{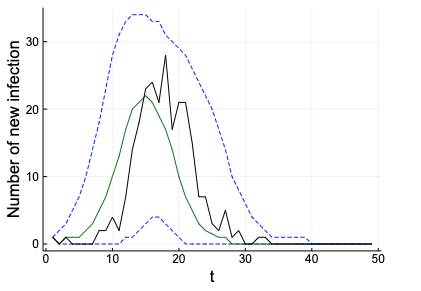}
         \caption{$\delta = (10, 20)$}
         \label{fig:three sin x}
     \end{subfigure}
     \hfill
     \begin{subfigure}[b]{0.32\textwidth}
         \centering
         \includegraphics[width=\textwidth]{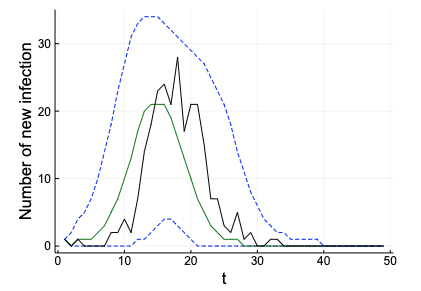}
         \caption{$\delta = (8,16)$}
         \label{fig:three sin x}
     \end{subfigure}
     \hfill
     \begin{subfigure}[b]{0.32\textwidth}
         \centering
         \includegraphics[width=\textwidth]{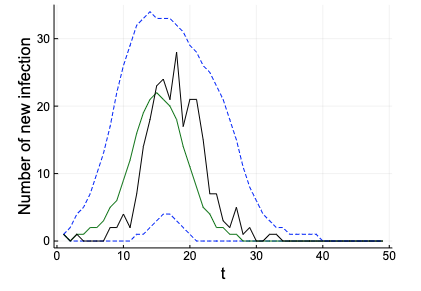}
         \caption{$\delta= (2,4,6)$}
         \label{fig:three sin x}
     \end{subfigure}
     \hfill
     \begin{subfigure}[b]{0.32\textwidth}
         \centering
         \includegraphics[width=\textwidth]{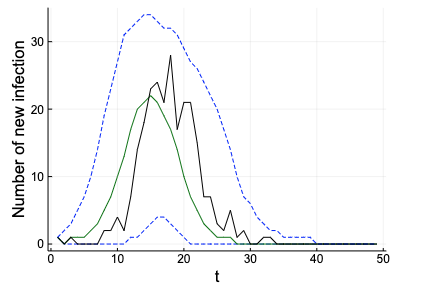}
         \caption{$\delta= (6,12,18)$}
         \label{fig:three sin x}
     \end{subfigure}
     \hfill
     \begin{subfigure}[b]{0.32\textwidth}
         \centering
         \includegraphics[width=\textwidth]{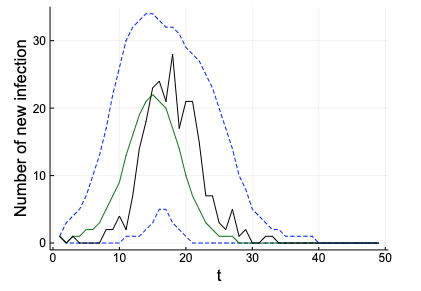}
         \caption{$\delta = (2,4,6,8)$}
         \label{fig:three sin x}
     \end{subfigure}
     \hfill
     \begin{subfigure}[b]{0.32\textwidth}
         \centering
         \includegraphics[width=\textwidth]{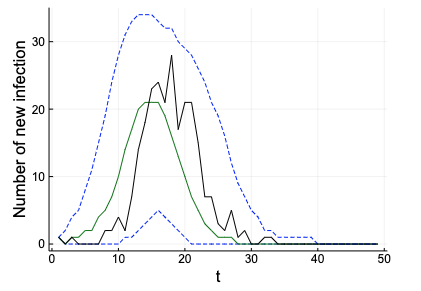}
         \caption{$\delta = (5,10,15,20) $}
         \label{fig:three sin x}
     \end{subfigure}
     \caption{\small{Posterior predictive distribution of the epidemic curve for one typical simulated epidemic for the true power-law and proposed semi-parametric models with piecewise constant kernel for different fixed change points. The black solid line represents the true epidemic curve, the green solid line represents the estimated median curve and the blue dotted lines represent the $95\%$ PI, respectively.}}\label{fig:1_fixed_PI}
\end{figure}

\begin{figure}[h!tpb]
     \centering
     \begin{subfigure}[b]{0.45\textwidth}
         \centering
         \includegraphics[width=\textwidth]{P_f.png}
         \caption{Parametric model}
         \label{fig:y equals x}
     \end{subfigure}
     \hfill
     \begin{subfigure}[b]{0.45\textwidth}
         \centering
         \includegraphics[width=\textwidth]{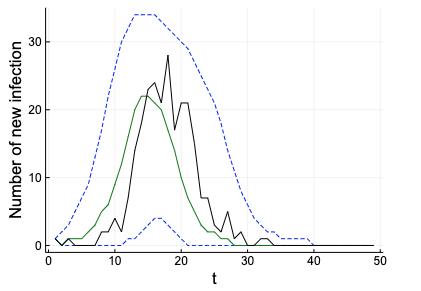}
         \caption{$\delta = 2.78$}
         \label{fig:three sin x}
     \end{subfigure}
     \hfill
     \begin{subfigure}[b]{0.45\textwidth}
         \centering
         \includegraphics[width=\textwidth]{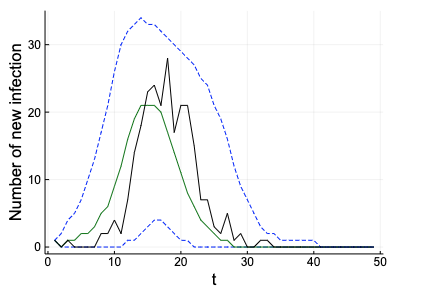}
         \caption{$\delta = (1.83,6.61)$}
         \label{fig:three sin x}
     \end{subfigure}
     
     \caption{Posterior predictive distribution of the epidemic curve for the true power-law and proposed semi-parametric models with piecewise constant kernel for different estimated change points. The black solid line represents the true epidemic curve, the green solid line represents the estimated median curve and the blue dotted lines represent the $95\%$ PI, respectively.}\label{fig:2_varying_PI}
\end{figure}

\begin{figure}[h!tpb]
    \centering
    \includegraphics[width=\textwidth]{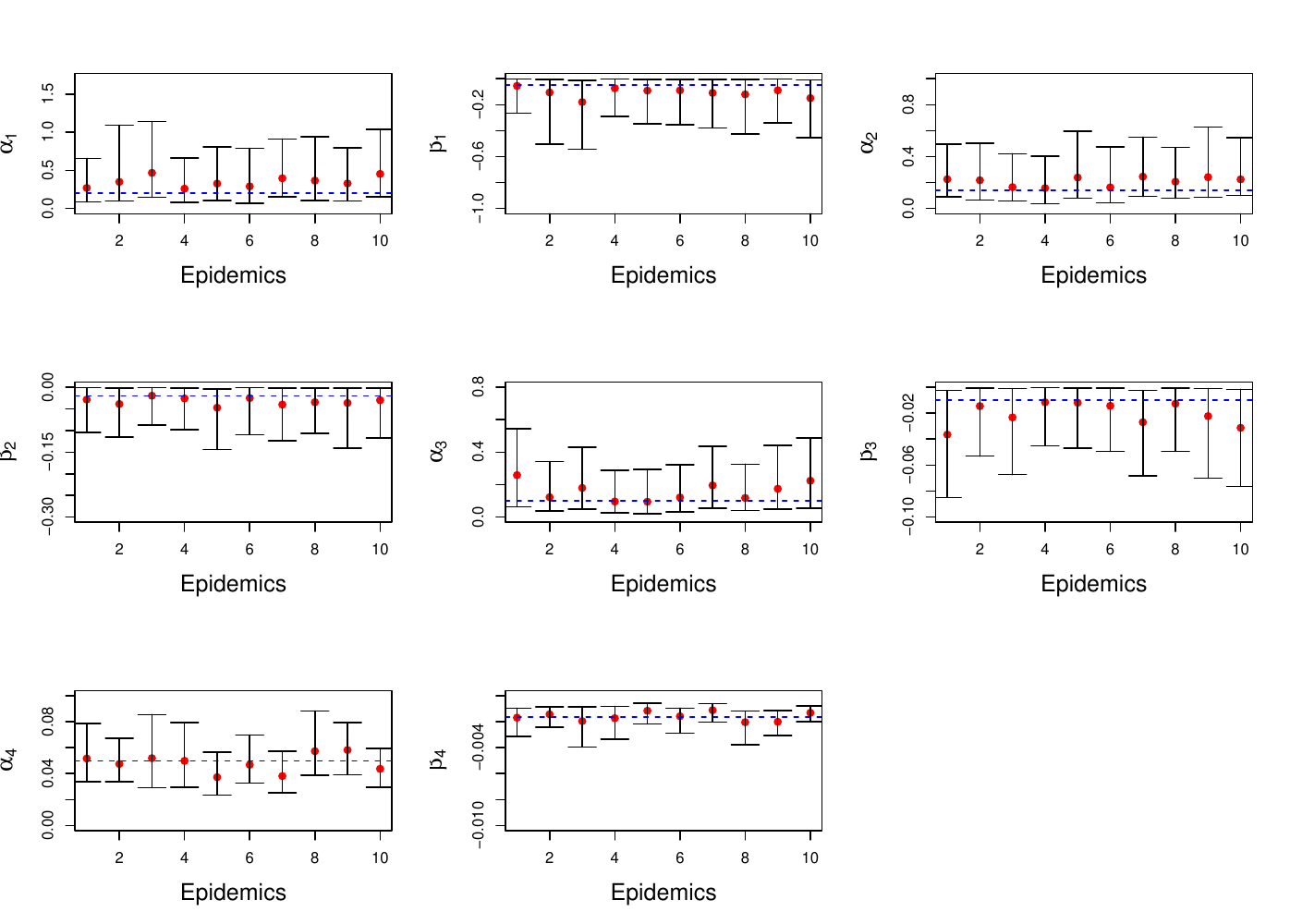}
     \caption{\small{Posterior medians (red points) and $95\%$ PIs for $\alpha_i,\beta_i$ where $i = 1,2,3,4$ for $10$ different simulated epidemics when the change points ($\delta$'s) are considered fixed. The true parameter values $\alpha_1 = 0.20, \beta_1 = -0.05, \alpha_2 = 0.14, \beta_2 = -0.02, \alpha_3 = 0.10, \beta_3 = -0.01, \alpha_4 = 0.0499, \beta_4 = -0.00165$  for three change points linear kernel models with fixed $\delta_1 = 2, \delta_2 = 4$, and $\delta_3 = 6$ are represented by the blue dashed line.} }\label{fig:L_f3_graphs}
\end{figure}

\begin{figure}[h!tpb]
     \centering
     \begin{subfigure}[b]{1\textwidth}
         \centering
         \includegraphics[width=0.71\textwidth]{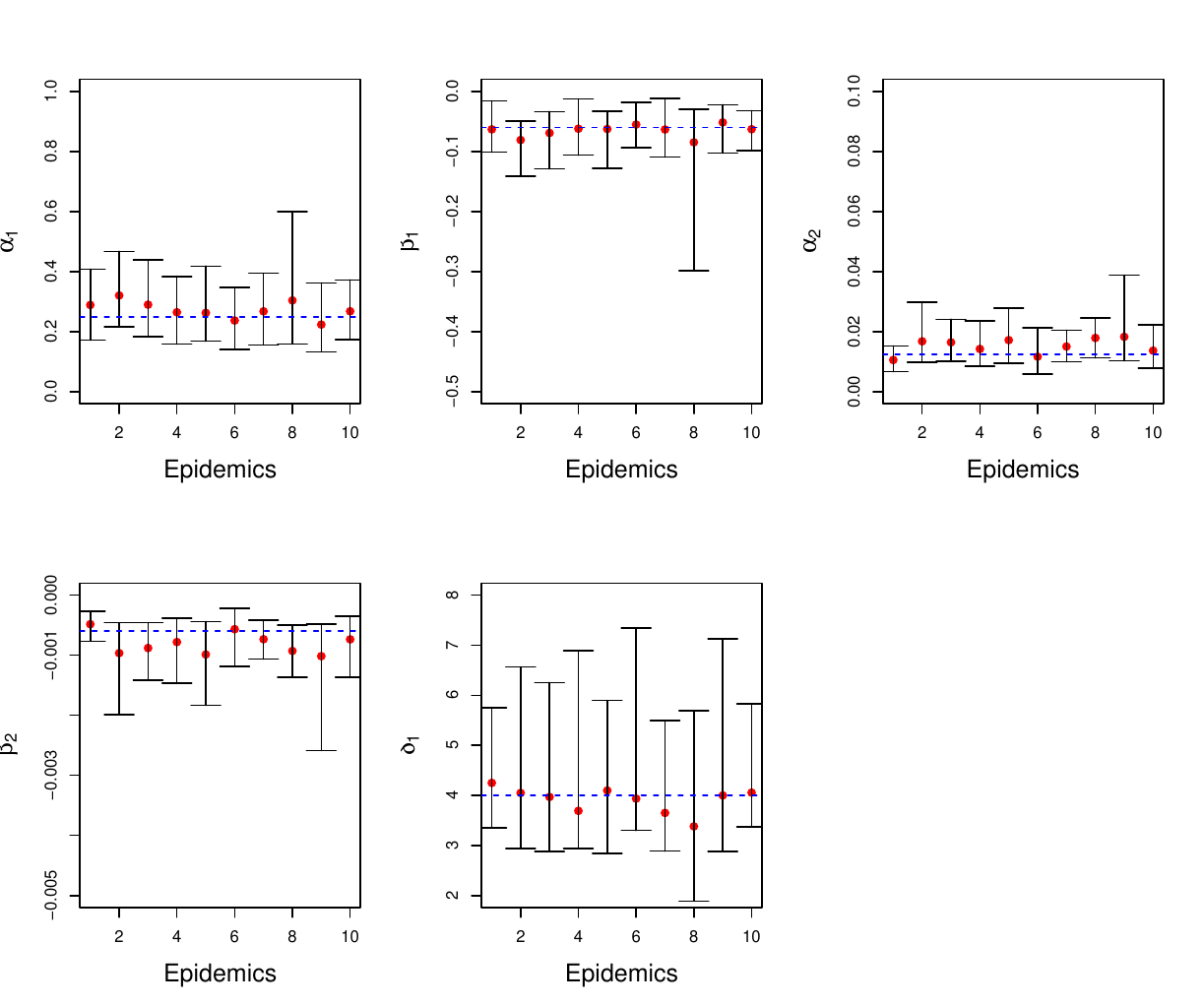}
         \caption{}
         \label{fig:y equals x}
     \end{subfigure}
     \vfill
     \begin{subfigure}[b]{1\textwidth}
         \centering
     \includegraphics[width=0.71\textwidth]{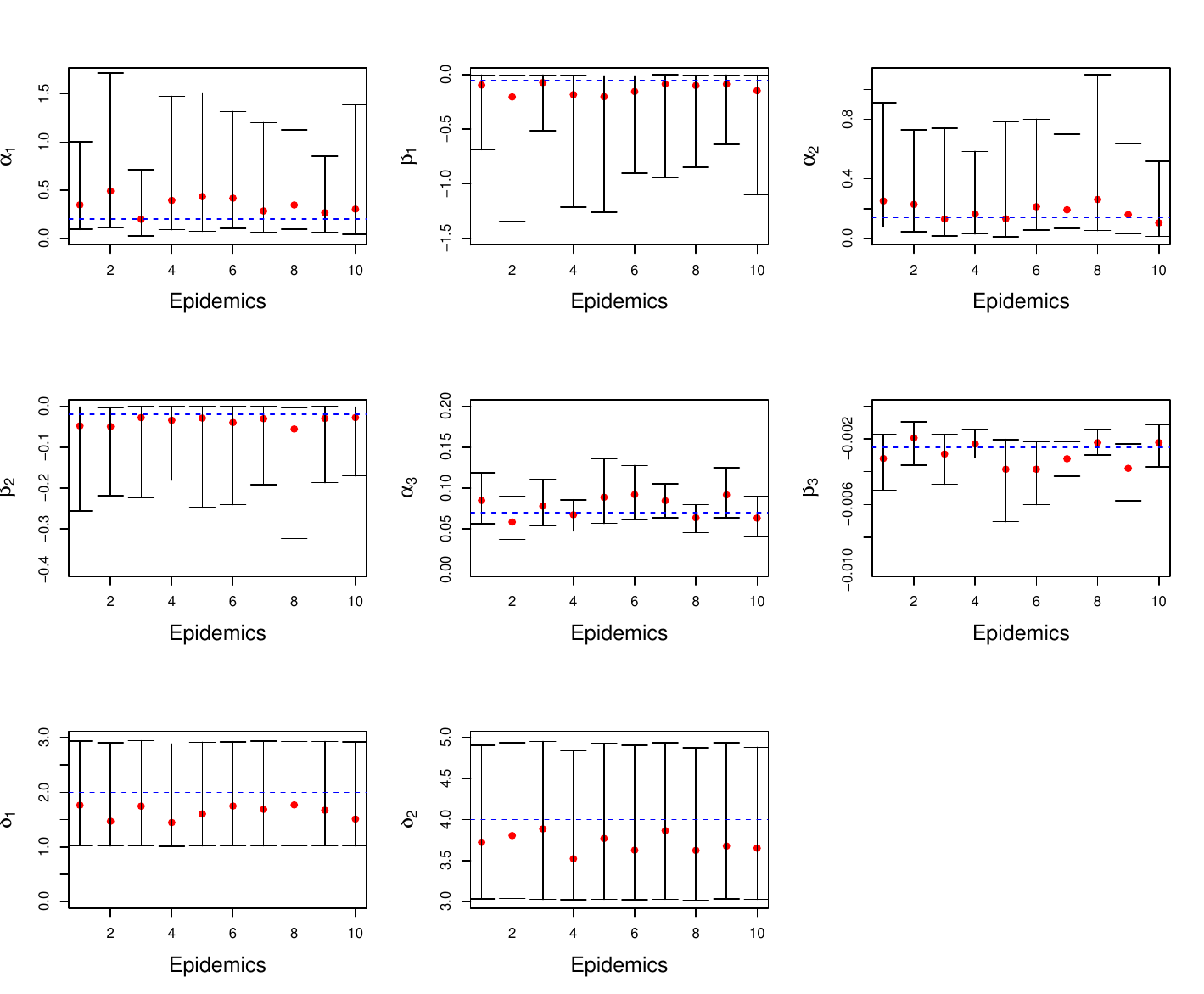}
         \caption{}
         \label{fig:three sin x}
     \end{subfigure}
     \caption{\small{Posterior medians (red points) and $95\%$ PIs for $\alpha_i,\beta_i$ where $i = 1,2,3$ for $10$ different simulated epidemics when the change points ($\delta$'s) are estimated. The true parameter values (a) $\alpha_1 = 0.25, \beta_1 = -0.06, \alpha_2 = 0.0124, \beta_2 = -0.0006,\delta  = 2$ for one change point linear kernel models, and (b) $\alpha_1 = 0.20, \beta_1 = -0.05, \alpha_2 = 0.14, \beta_2 = -0.02, \alpha_3 = 0.07, \beta_3 = -0.0025,\delta_1 = 2, \delta_2 = 4 $  for two change points linear kernel models are represented by the blue dashed line. } }\label{fig:Lv_graphs}
\end{figure}

\begin{figure}[h!tpb]
      \centering
	   \begin{subfigure}{0.325\linewidth}
		\includegraphics[width=\linewidth]{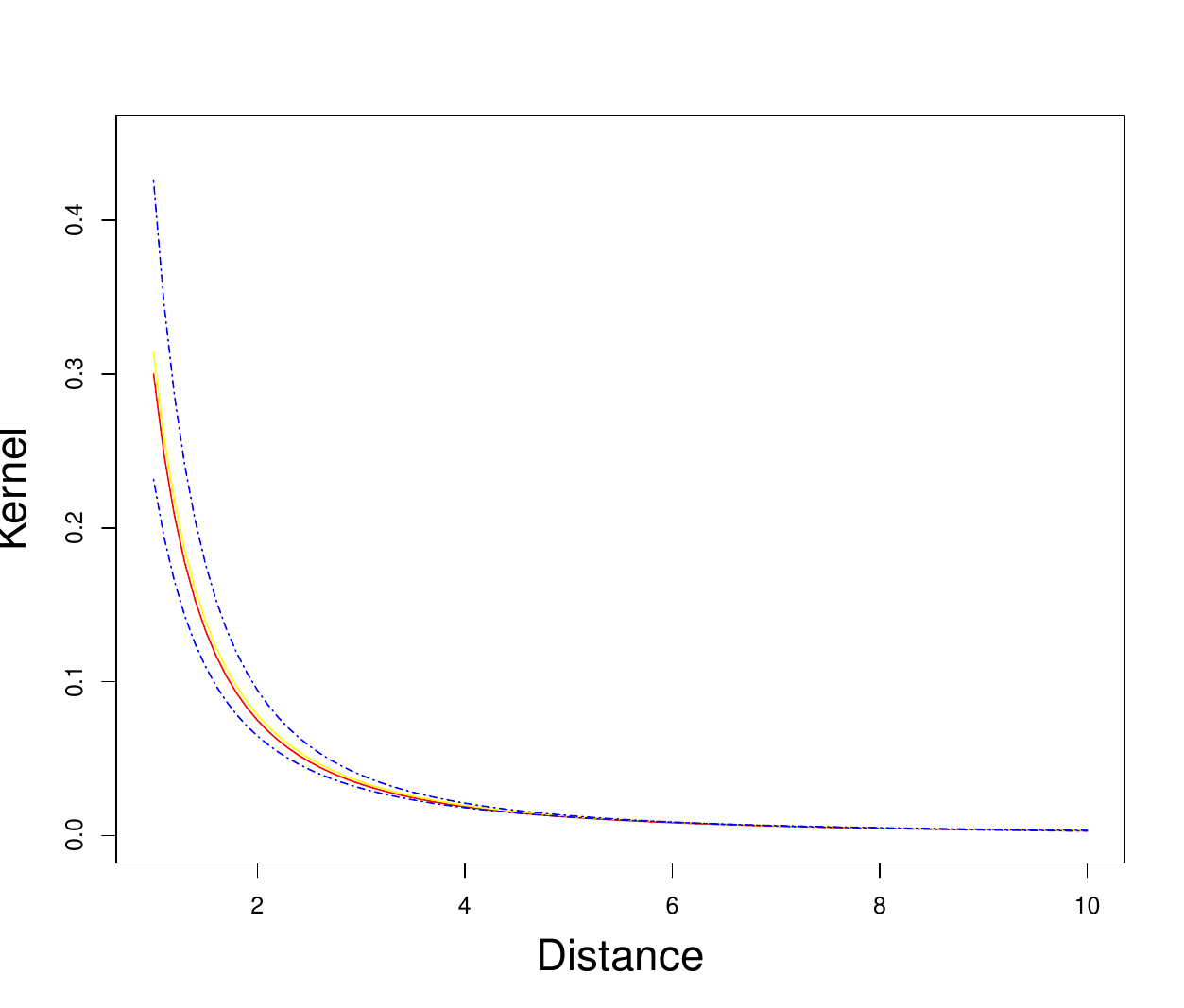}
		\caption{}
		\label{fig:subfig1}
	   \end{subfigure}
	   \begin{subfigure}{0.325\linewidth}
		\includegraphics[width=\linewidth]{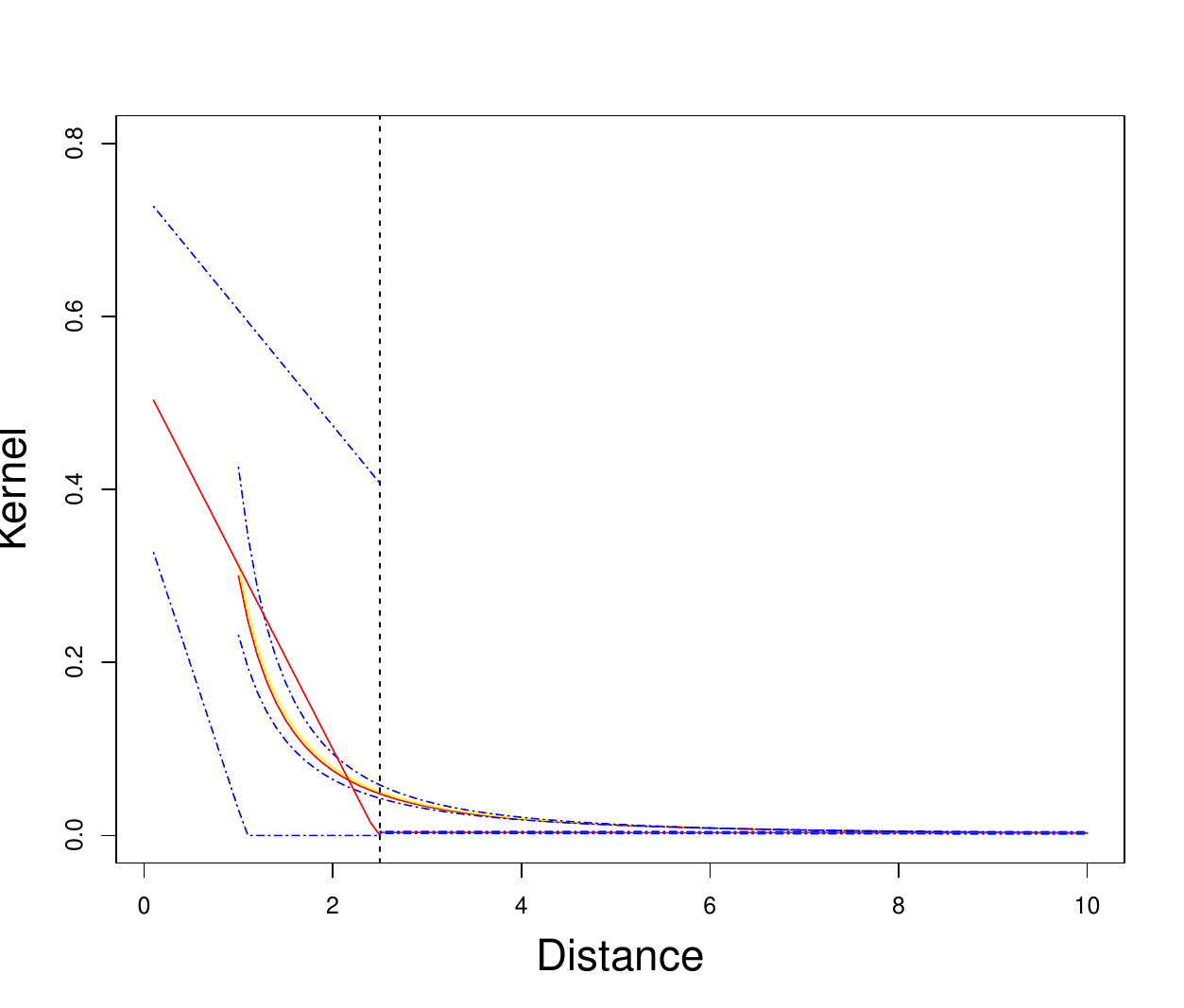}
		\caption{$\delta = 2.5, D = 0.037 $}
		\label{fig:subfig2}
	    \end{subfigure}
        \begin{subfigure}{0.325\linewidth}
		\includegraphics[width=\linewidth]{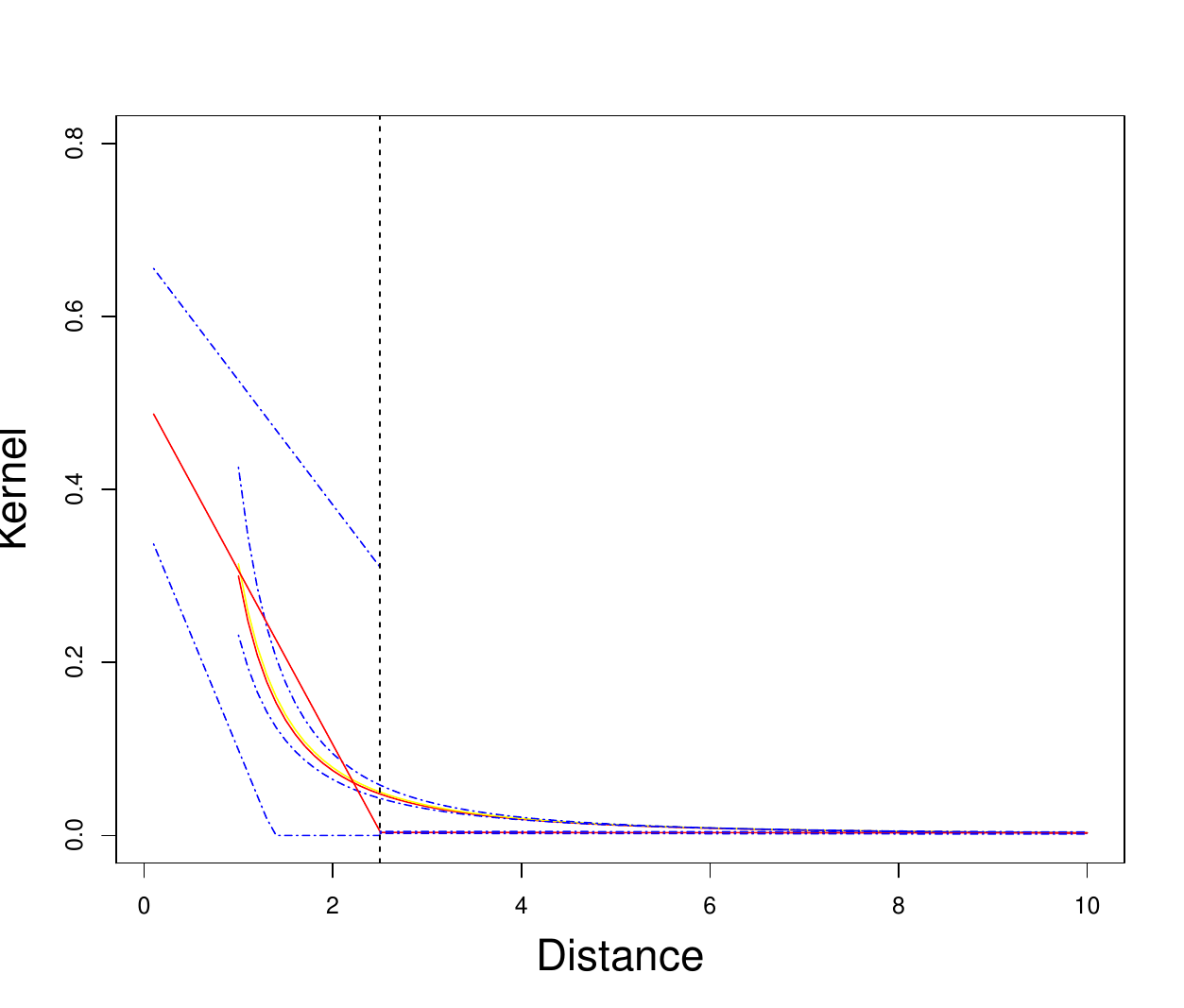}
		\caption{$\delta = 2.5,D= 0.003$}
		\label{fig:subfig2}
	    \end{subfigure}
	\vfill
	     \begin{subfigure}{0.325\linewidth}
		 \includegraphics[width=\linewidth]{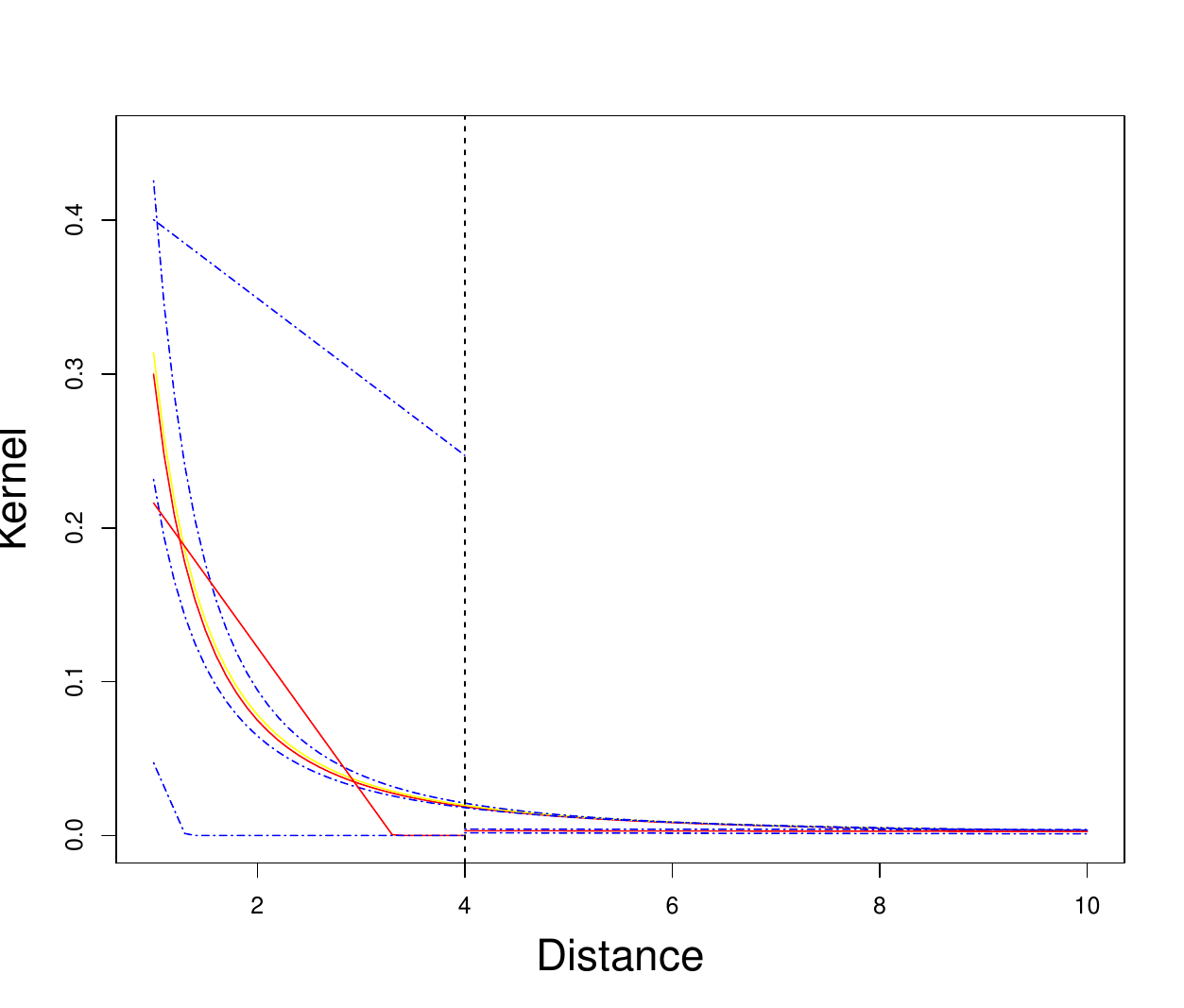}
		 \caption{$\delta = 4,D = 0.037 $}
		 \label{fig:subfig3}
	      \end{subfigure}
	       \begin{subfigure}{0.325\linewidth}
		  \includegraphics[width=\linewidth]{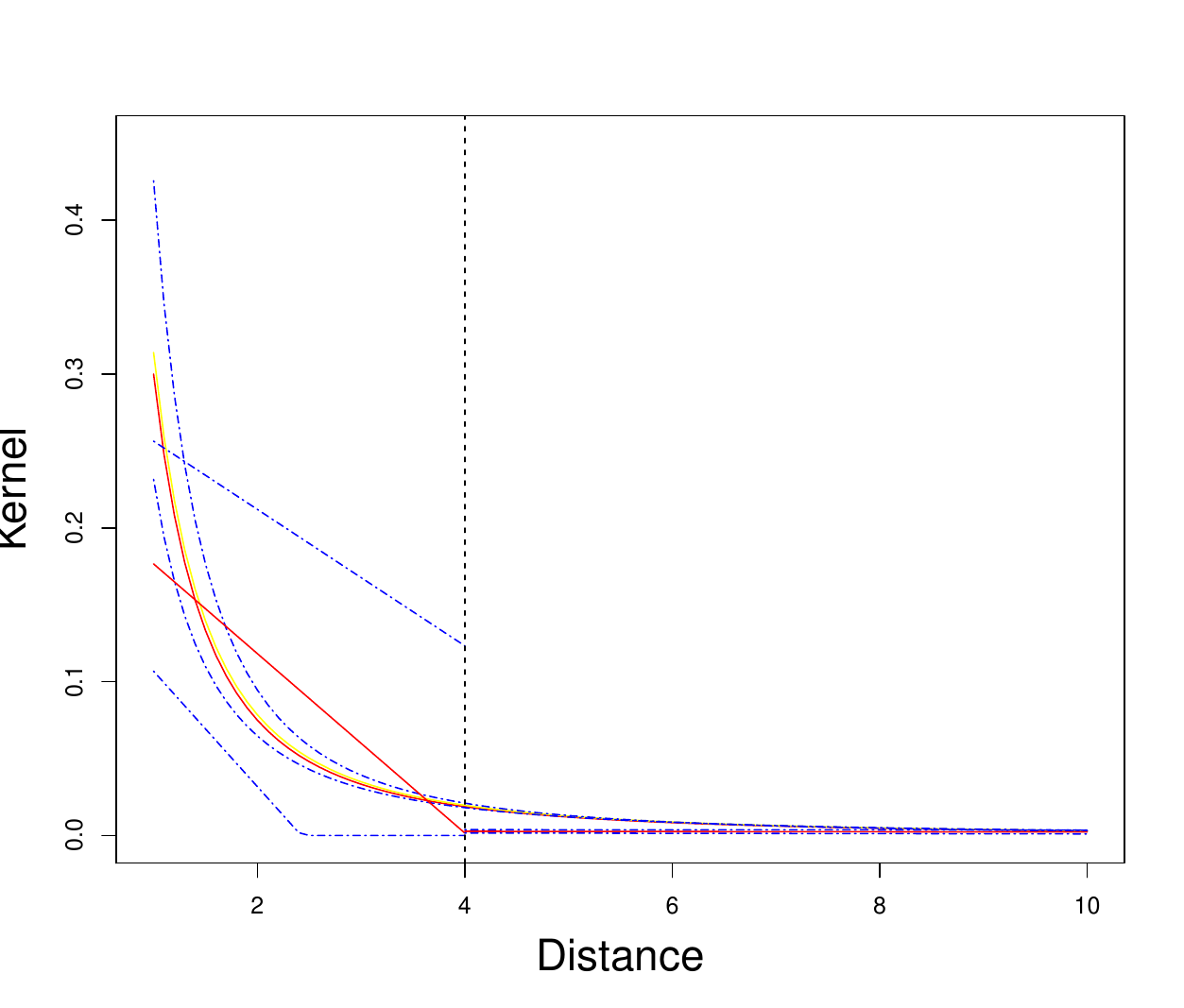}
		  \caption{$\delta = 4,D = 0.003$}
		  \label{fig:subfig4}
	       \end{subfigure}
        \begin{subfigure}{0.325\linewidth}
		  \includegraphics[width=\linewidth]{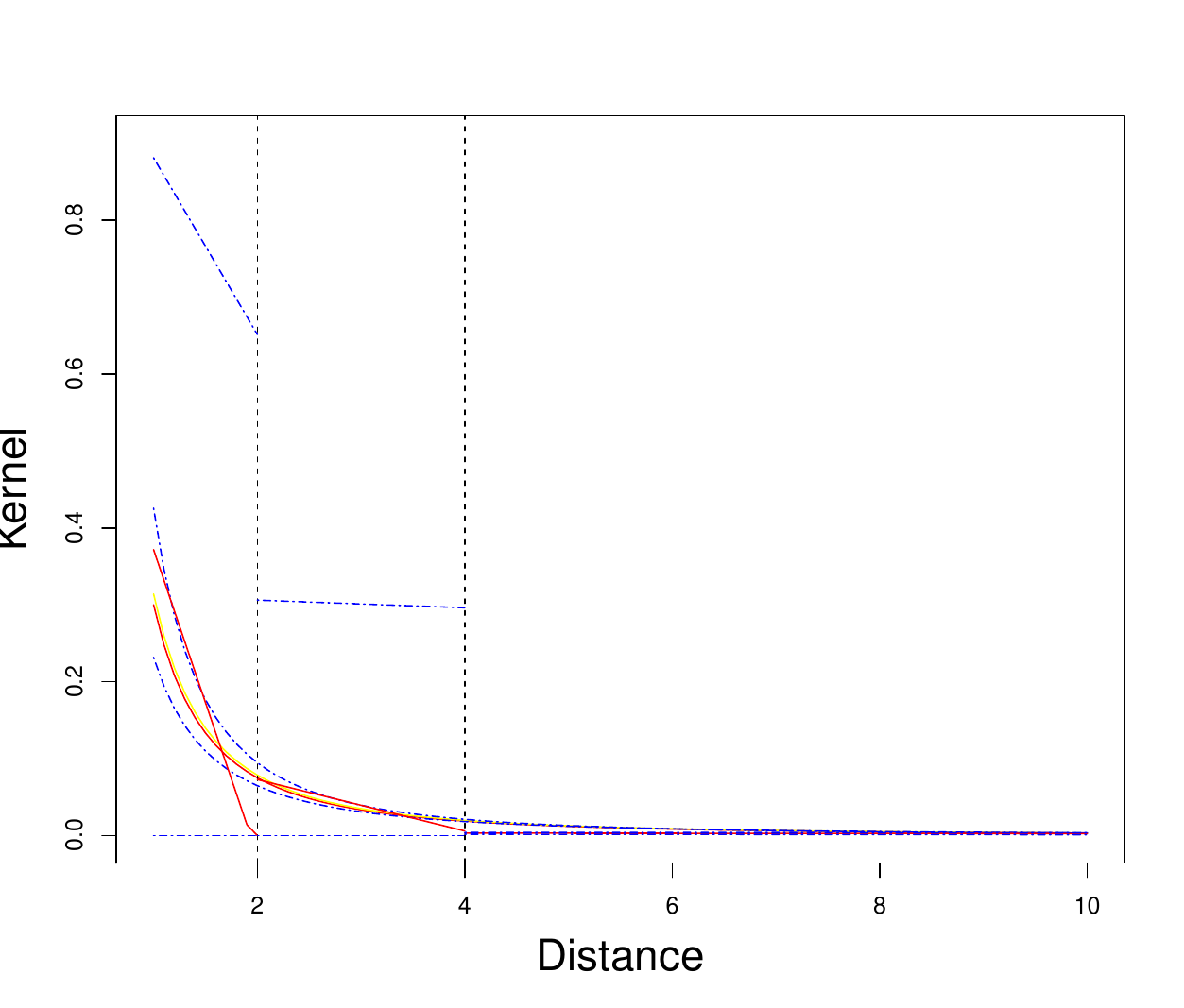}
		  \caption{$\delta = (2,4),D =(0.047,0.037)$}
		  \label{fig:subfig4}
	       \end{subfigure}
        \vfill
        \begin{subfigure}{0.325\linewidth}
		  \includegraphics[width=\linewidth]{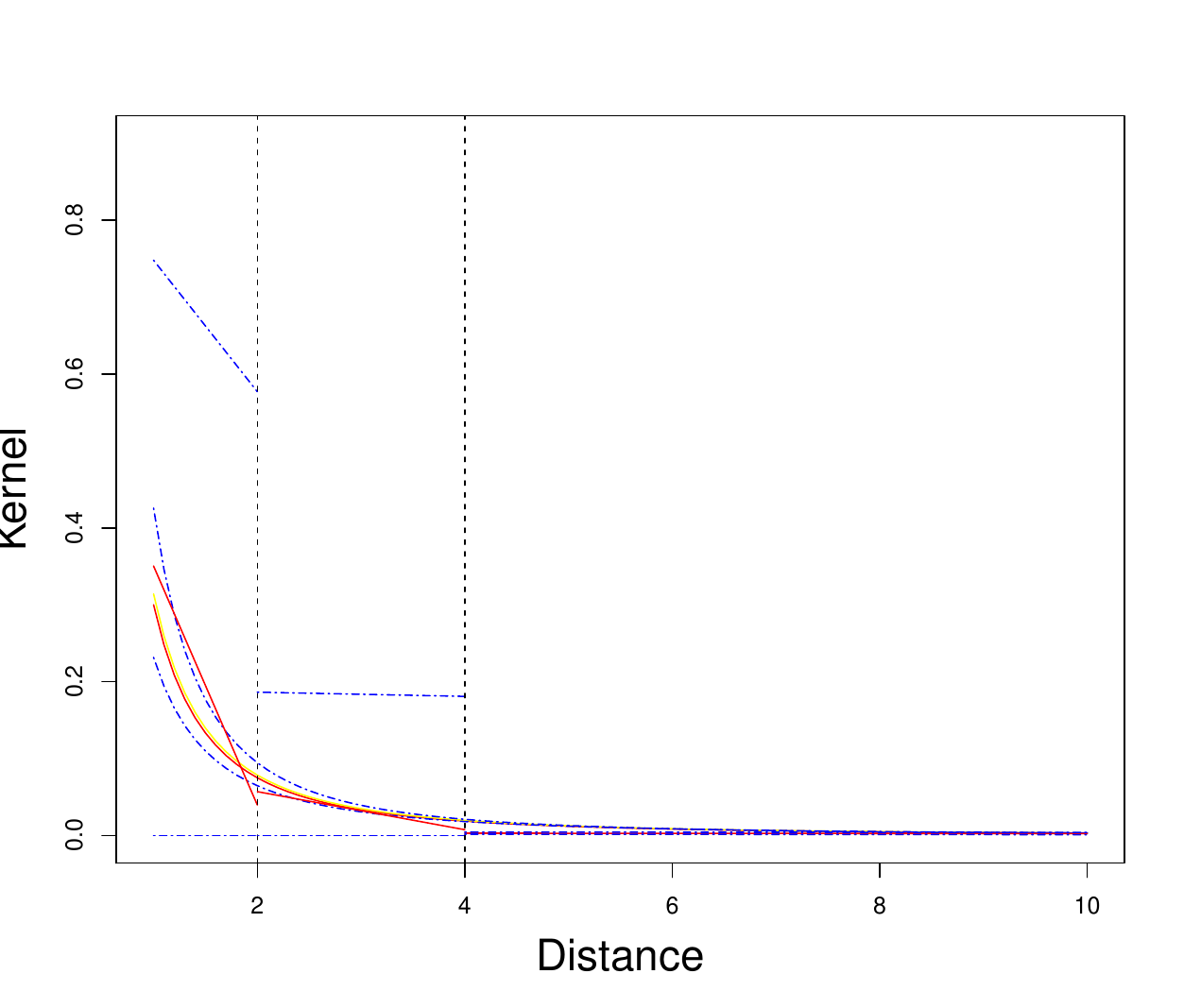}
		  \caption{$\delta = (2,4),D =(0.01,0.01)$ }
		  \label{fig:subfig4}
	       \end{subfigure}
        \begin{subfigure}{0.33\linewidth}
		  \includegraphics[width=\linewidth]{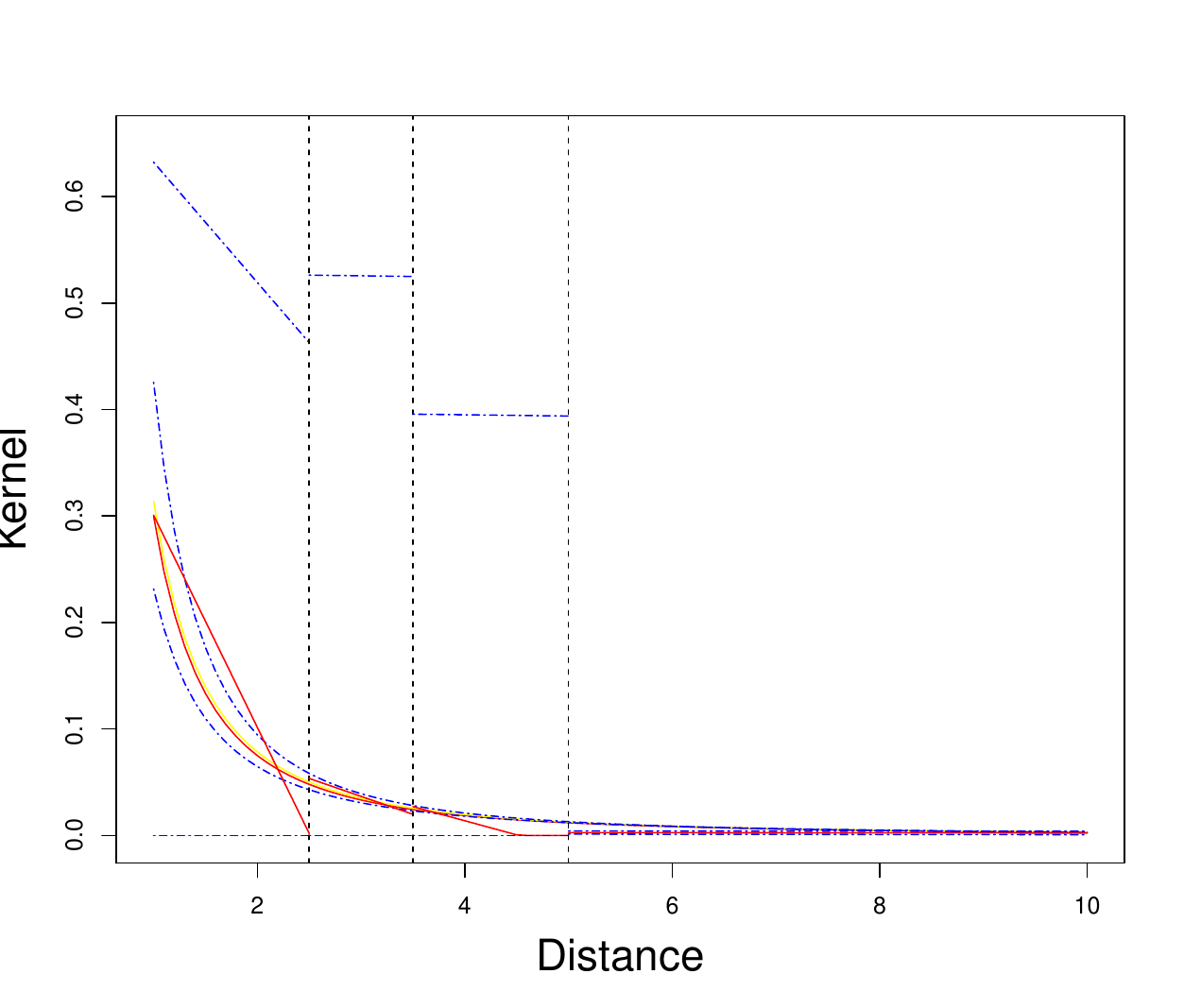}
		  \caption{$\delta=(2.5,3.5,5)$,
                $D=(0.05,0.05,0.05)$}
		  \label{fig:subfig4}
	       \end{subfigure}
        \begin{subfigure}{0.33\linewidth}
		  \includegraphics[width=\linewidth]{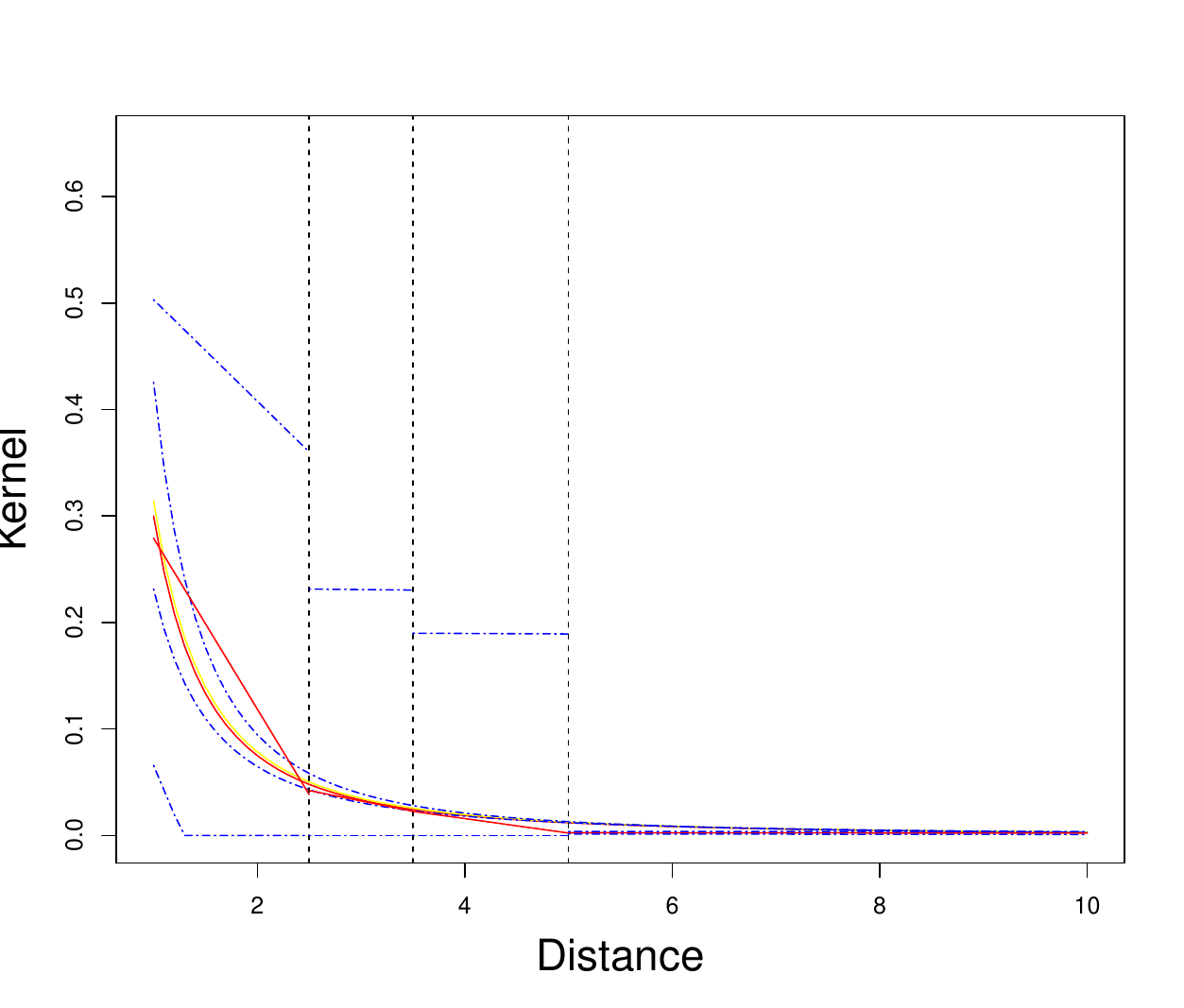}
		  \caption{$\delta=(2.5,3.5,5)$,
                $D=(0.01,0.01,0.01)$}
		  \label{fig:subfig4}
	       \end{subfigure}
        
	\caption{\small{True kernel,  fitted kernel under the  posterior median and $95\%$ PIs for (a) the power-law model are represented by red (solid), yellow (solid) and blue (dashed) lines, respectively. Linear lines under the posterior medians and respective $95\%$ PIs for models (b-i) with different change points ($\delta$) (black dashed line) and smoothing parameters ($D$) are represented by red  and blue dashed lines, respectively.}}\label{fig:linear_smoothing}
\end{figure}

\begin{figure}[h!tpb]
     \begin{subfigure}[b]{0.32\textwidth}
         \centering
         \includegraphics[width=\textwidth]{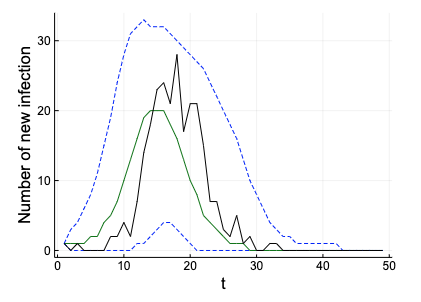}
         \caption{Parametric model}
         \label{fig:linear1}
     \end{subfigure}
     \hfill
     \begin{subfigure}[b]{0.32\textwidth}
         \centering
         \includegraphics[width=\textwidth]{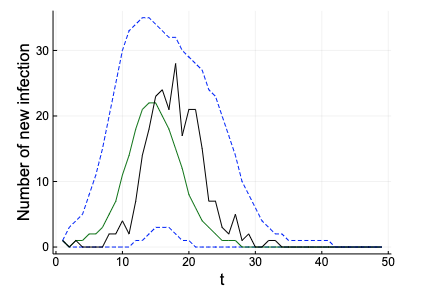}
         \caption{$\delta = 2.5, D = 0.037$}
         \label{fig:linear2}
     \end{subfigure}
     \hfill
     \begin{subfigure}[b]{0.32\textwidth}
         \centering
         \includegraphics[width=\textwidth]{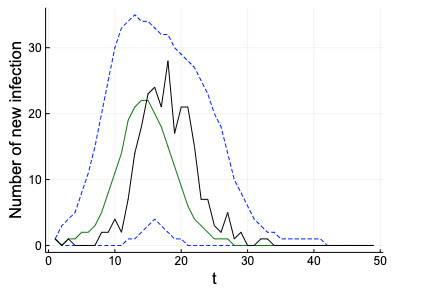}
         \caption{$\delta = 2.5, D= 0.003$}
         \label{fig:linear3}
     \end{subfigure}
     \hfill
     \begin{subfigure}[b]{0.32\textwidth}
         \centering
         \includegraphics[width=\textwidth]{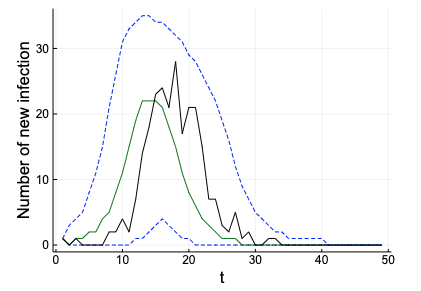}
         \caption{$\delta = 4, D = 0.037 $}
         \label{fig:linear4}
     \end{subfigure}
    \hfill
     \begin{subfigure}[b]{0.32\textwidth}
         \centering
         \includegraphics[width=\textwidth]{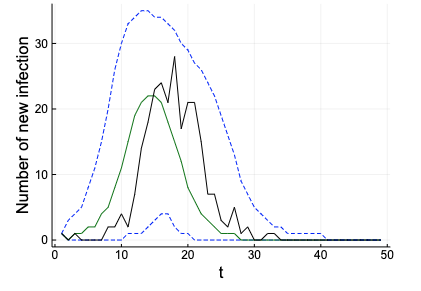}
         \caption{$\delta = 4, D = 0.003$}
         \label{fig:linear5}
     \end{subfigure}
     \hfill
     \begin{subfigure}[b]{0.32\textwidth}
         \centering
         \includegraphics[width=\textwidth]{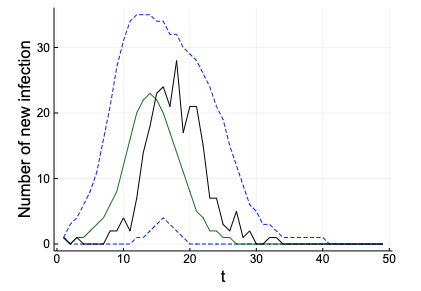}
         \caption{$\delta = (2,4), D = (0.047,0.037)$}
         \label{fig:linear6}
     \end{subfigure}
     \hfill
     \begin{subfigure}[b]{0.32\textwidth}
         \centering
         \includegraphics[width=\textwidth]{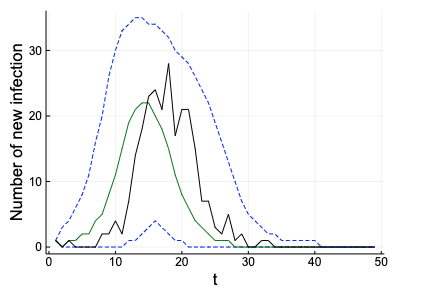}
         \caption{$\delta = (2,4), D = (0.01,0.01)$}
         \label{fig:linear7}
     \end{subfigure}
     \hfill
     \begin{subfigure}[b]{0.32\textwidth}
         \centering
         \includegraphics[width=\textwidth]{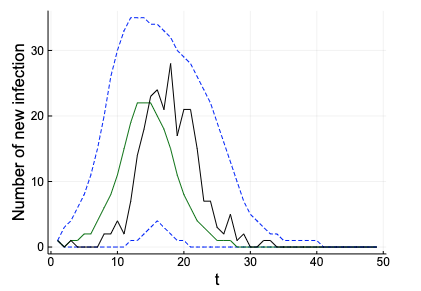}
         \caption{$\delta = (2.5,3.5,5), D = (0.05,0.05,0.05)$}
         \label{fig:linear8}
     \end{subfigure}
     \hfill
     \begin{subfigure}[b]{0.32\textwidth}
         \centering
         \includegraphics[width=\textwidth]{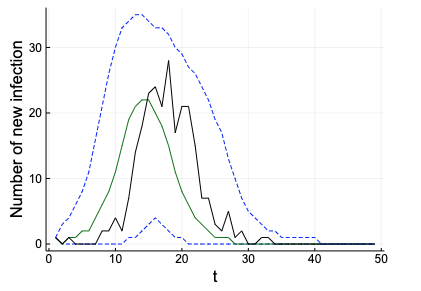}
         \caption{$\delta =(2.5,3.5,5),D = (0.01,0.01,0.01)$}
         \label{fig:linear9}
     \end{subfigure}
     \caption{\small{Posterior predictive distribution of the epidemic curve for one typical simulated epidemic for the true power-law and proposed semi-parametric models with piecewise linear kernel for different fixed change points along with smoothing prior with different scale parameters. The black solid line represents the true epidemic curve, the green solid line represents the estimated median curve and the blue dotted lines represent the $95\%$ PI, respectively.}}\label{fig:L_fixed_PI}
\end{figure}

\begin{figure}[h!tpb]
      \centering
	   \begin{subfigure}{0.24\linewidth}
		\includegraphics[width=\linewidth]{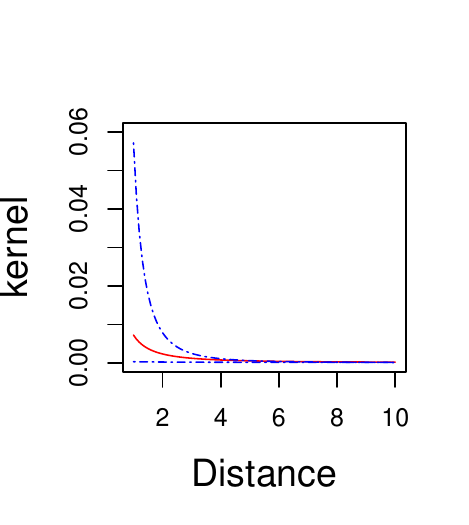}
		\caption{}
		\label{fig:subfig1}
	   \end{subfigure}
	   \begin{subfigure}{0.24\linewidth}
		\includegraphics[width=\linewidth]{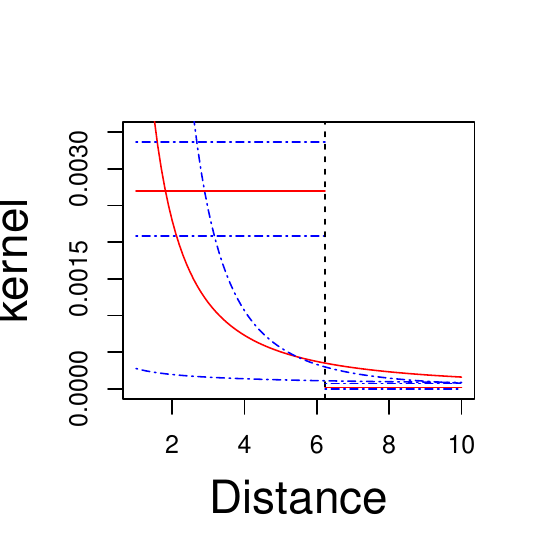}
		\caption{$\delta = 6.19$}
		\label{fig:subfig2}
	    \end{subfigure}
        \begin{subfigure}{0.24\linewidth}
		\includegraphics[width=\linewidth]{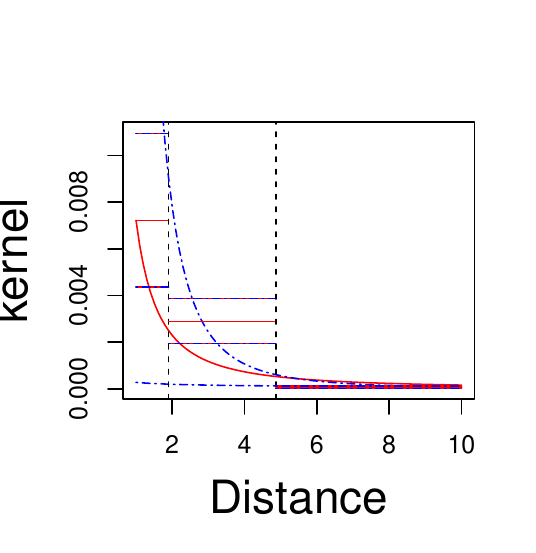}
		\caption{$\delta = (1.92,4.85)$}
		\label{fig:subfig2}
	    \end{subfigure}
	     \begin{subfigure}{0.24\linewidth}
		 \includegraphics[width=\linewidth]{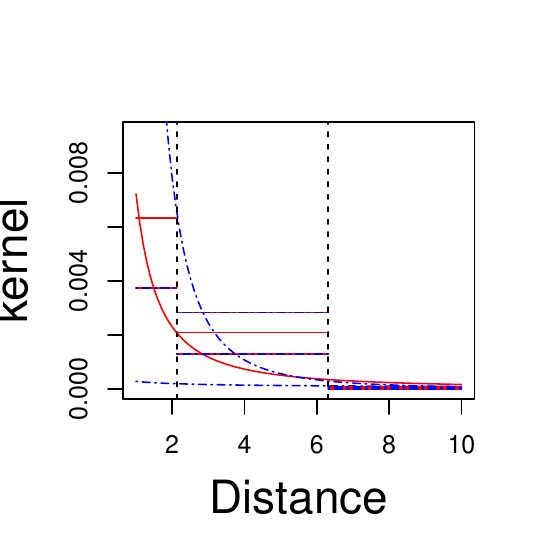}
		 \caption{$\delta = (2.13,6.18))$}
		 \label{fig:subfig3}
	      \end{subfigure}
       \caption{\small{Fitted kernel under the posterior median and $95\%$ PIs for (a) the power-law kernel model are represented by red (solid), and blue dashed lines, respectively. posterior medians and $95\%$ PIs for different estimated change point ($\delta$) models (b-d) with piecewise constant kernel which fitted to FMD data are represented by red and blue dashed lines,respectively.}}\label{fig:FMD_para_varying}
\end{figure}

\begin{figure}[h!tpb]
      \centering
	   \begin{subfigure}{0.325\linewidth}
		\includegraphics[width=\linewidth]{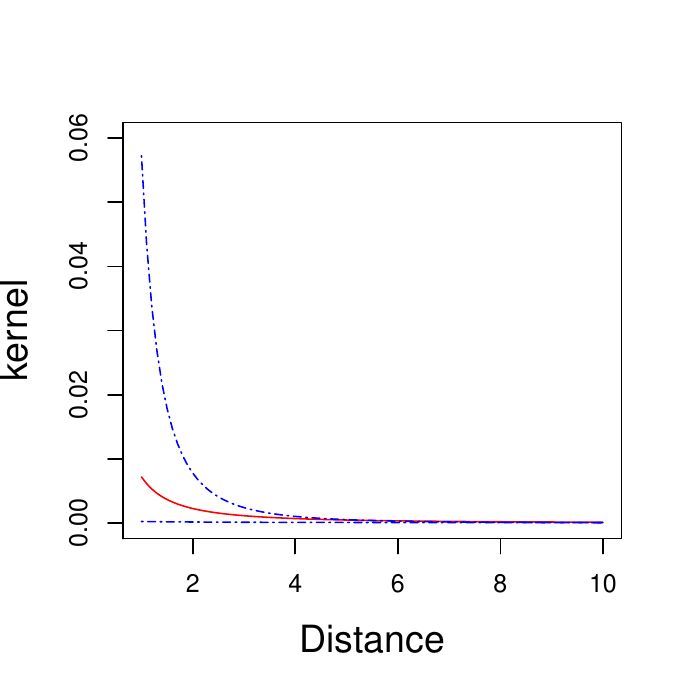}
		\caption{}
		\label{fig:subfig1}
	   \end{subfigure}
	   \begin{subfigure}{0.325\linewidth}
		\includegraphics[width=\linewidth]{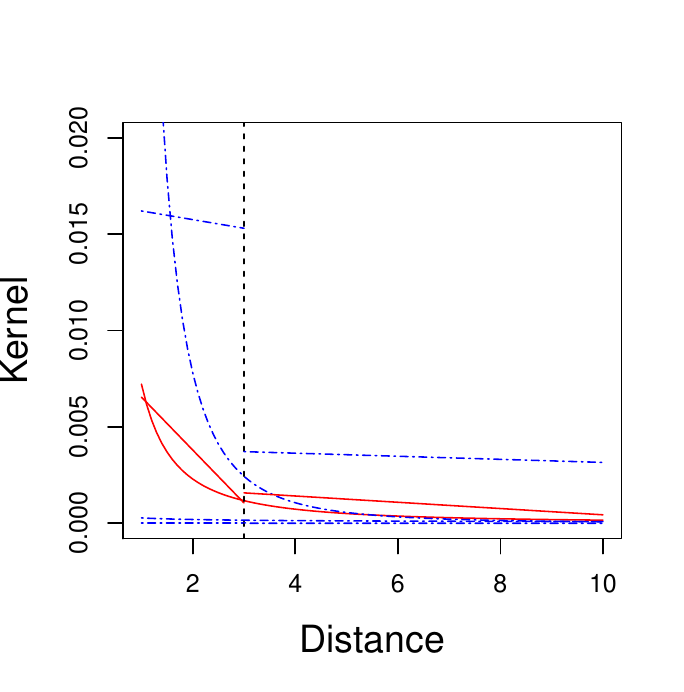}
		\caption{$\delta = 3, D = 0.10 $}
		\label{fig:subfig2}
	    \end{subfigure}
        \begin{subfigure}{0.325\linewidth}
		\includegraphics[width=\linewidth]{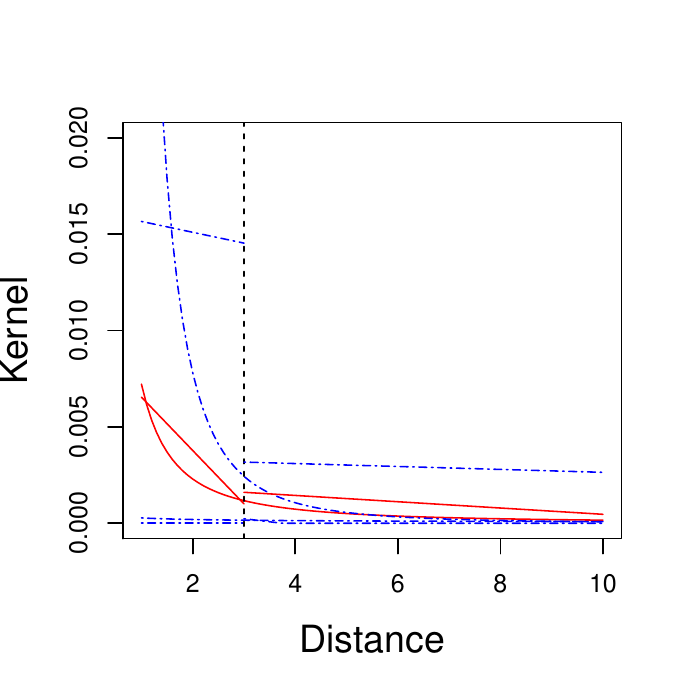}
		\caption{$\delta = 3, D= 0.04$}
		\label{fig:subfig2}
	    \end{subfigure}

	\caption{\small{Fitted kernel,  and $95\%$ PIs for (a) the power-law kernel model are represented by red (solid), and blue dashed lines, respectively. Linear lines under the posterior medians and respective $95\%$ PIs for models (b-c) with different change points ($\delta$) (black dashed line) and smoothing parameters ($D$) which fitted to FMD data are represented by red and blue dashed lines, respectively.}}\label{fig:FMD_f_linear_smoothing}
\end{figure}

 \newpage
 \clearpage

    \addcontentsline{toc}{section}{\refname}

    \bibliographystyle{unsrtnat}
    \bibliography{bibliography.bib}

	\newpage

\end{document}